\documentclass[aps,pra,twocolumn,groupedaddress,floatfix,superscriptaddress]{revtex4-1}

\usepackage{amsmath}
\usepackage{amssymb}
\usepackage{graphicx}
\usepackage{float}
\usepackage[english]{babel}
\usepackage{dsfont}
\usepackage{epstopdf}
\usepackage{soul}
\usepackage{color}
\usepackage{braket}
\usepackage[colorlinks=true,
            linkcolor=magenta,
            urlcolor=blue,
            citecolor=blue]{hyperref}

\usepackage[normalem]{ulem}

\def\be{\begin{equation}}
\def\ee{\end{equation}} 
\def\bea{\begin{eqnarray}}
\def\eea{\end{eqnarray}} 
\def\ba{\begin{array}} 
\def\ea{\end{array}}

\def\om{\omega}

\def\b{\mathbf}
\def\bs{\boldsymbol}

\def\ra{\rightarrow}

\def\ua{\uparrow}
\def\da{\downarrow}

\emergencystretch=\maxdimen
\begin{document}
\title{Strain and pseudo-magnetic fields in optical lattices from density-assisted tunneling}
\author{M. Jamotte}
\email{mjamotte@ulb.be}
\affiliation{Center for Nonlinear Phenomena and Complex Systems, Universit\'e Libre de Bruxelles, CP 231, Campus Plaine, B-1050 Brussels, Belgium}
\author{N. Goldman}
\email{ngoldman@ulb.be}
\affiliation{Center for Nonlinear Phenomena and Complex Systems, Universit\'e Libre de Bruxelles, CP 231, Campus Plaine, B-1050 Brussels, Belgium}
\author{M. Di Liberto}
\email{mar.diliberto@gmail.com}
\affiliation{Center for Nonlinear Phenomena and Complex Systems, Universit\'e Libre de Bruxelles, CP 231, Campus Plaine, B-1050 Brussels, Belgium}
\affiliation{Institute for Quantum Optics and Quantum Information of the Austrian Academy of Sciences, Innsbruck, Austria}

\begin{abstract}

Applying time-periodic modulations is routinely used to control and design synthetic matter in quantum-engineered settings. In lattice systems, this approach is explored to engineer band structures with non-trivial topological properties, but also to generate exotic interaction processes. A prime example is density-assisted tunneling, by which the hopping amplitude of a particle between neighboring sites explicitly depends on their respective occupations. Here, we show how density-assisted tunneling can be tailored in view of simulating the effects of strain in synthetic graphene-type systems. Specifically, we consider a mixture of two atomic species on a honeycomb optical lattice: one species forms a Bose-Einstein condensate in an anisotropic harmonic trap, whose inhomogeneous density profile induces an effective uniaxial strain for the second species through density-assisted tunneling processes. In direct analogy with strained graphene, the second species experiences a pseudo magnetic field, hence exhibiting relativistic Landau levels and the valley Hall effect. Our proposed scheme introduces a unique platform for the investigation of strain-induced gauge fields and their possible interplay with quantum fluctuations and collective excitations.

\end{abstract}

\maketitle

\section{Introduction}
\label{sec:intro}

The rise of cold atoms in optical lattices as a versatile platform to study quantum phases \cite{Jaksch2005, Bloch_rmp, Bloch2012, Schafer2020} has led to the realization and investigation of rich physical models. Building on the milestone implementation of the Bose-Hubbard model~\cite{Greiner2002}, a model of bosonic particles on a lattice with onsite (contact) interactions~\cite{Jaksch1998}, a variety of opportunities have become available. For instance, the realization of the Fermi-Hubbard model with cold atoms~\cite{Kohl2005,Schneider2008,Boll2016, Mazurenko2017, Brown2019} represents a promising route to unveil the microscopic origin of high-temperature superconductivity~\cite{Hofstetter2002}. More recently, tunable long-range interactions have been introduced in atomic lattice systems~\cite{dePaz2013,yan2013observation,Baier2016}, as well as more exotic features~\cite{Dutta2015}, such as SU(N)-symmetric interactions~\cite{Taie2012,Zhang2014}, and density-assisted tunneling~\cite{Gong2009,Rapp2012, DiLiberto2014,Luhmann2014, Meinert2016, Gorg2018}. 


Gauge fields are at the core of remarkable phenomena in condensed matter, as was exemplified by the discovery of the quantum Hall effects and topological materials~\cite{Kane2010,Qi2011,armitage2018weyl}. 
These exciting topics have become accessible in ultracold gases through the design of synthetic gauge fields~\cite{Dalibard2011,goldman2014light,Goldman2016a,aidelsburger2018artificial,Cooper2019}. One of the key methods to realize synthetic gauge potentials in quantum-engineered systems consists in driving the system periodically in time \cite{Goldman2014,Bukov2015,Eckardt2017}, a general scheme also known as \emph{Floquet engineering}~\cite{Rudner2020}; in this driven-lattice context, tunneling matrix elements acquire well-designed complex phase factors, hence mimicking the Aharonov-Bohm effect caused by an external magnetic field. These Floquet schemes have been applied to generate the Harper-Hofstadter~\cite{Harper1955, Hofstadter1976, Jaksch2003, Aidelsburger2013, Miyake2013,Aidelsburger2015,Greiner2017} and Haldane-type \cite{Haldane1988, Jotzu2014,Wu2016,Asteria2019} models in optical lattices.


\begin{figure*}[!t]
	\center
	\includegraphics[width=\textwidth]{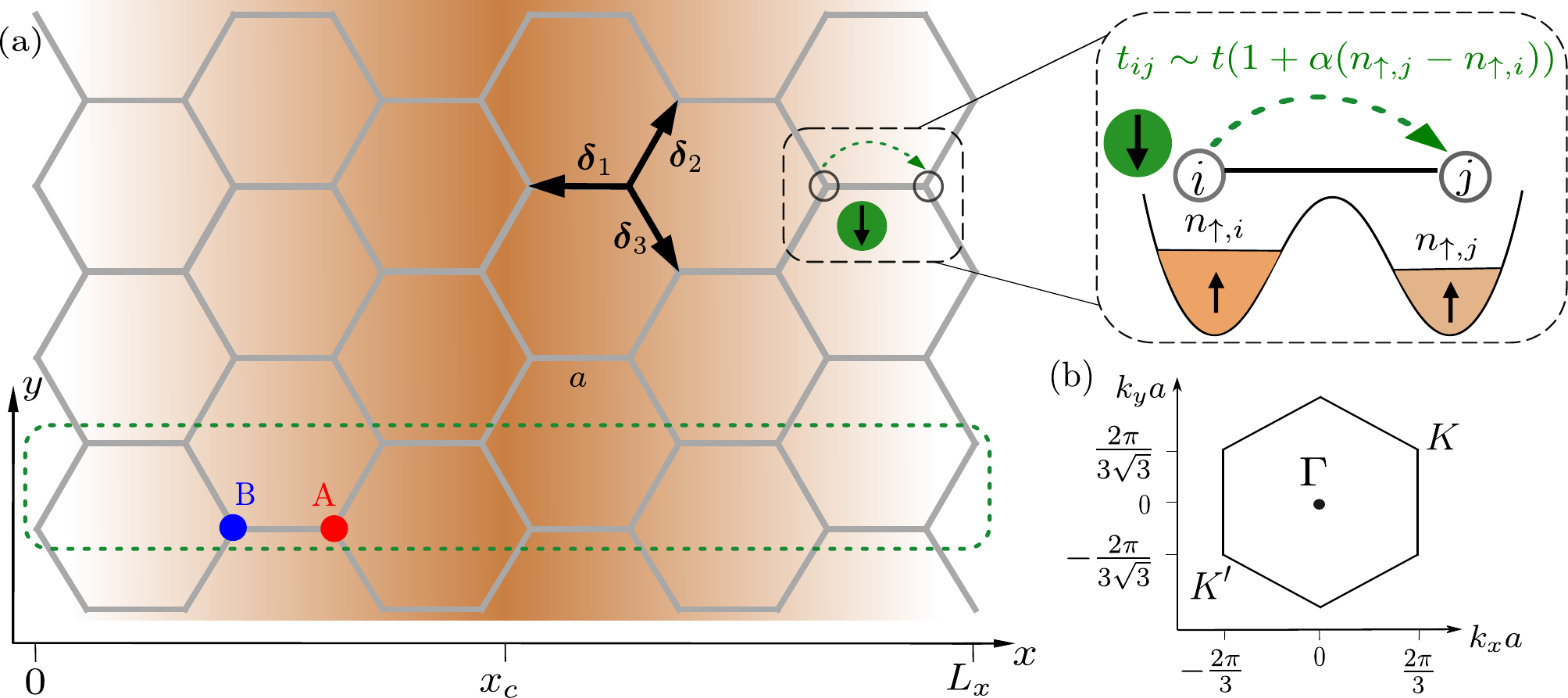}
	\caption{(a) Representation of the density-assisted hopping model on the honeycomb lattice. The shaded area depicts the inhomogeneous BEC cloud. In the zoomed panel, we depict the hopping process of the $\da$ atoms between two neighboring sites occupied by a different amount of $\ua$ atoms. The unit cell of the lattice where we assume translational invariance in the $y$ direction is shown by the green dotted line with zigzag terminations. (b) First Brillouin zone of the honeycomb lattice where Dirac points at $\mathbf K = (2\pi/3a, 2\pi/3\sqrt{3}a)$ and $\mathbf K' = -\mathbf K$ are indicated.}
	\label{fig:Fig_1}
\end{figure*}

An exciting scenario, which has become more concrete and realistic over the last few years, concerns the realization of dynamical gauge fields in cold gases, namely, engineered gauge fields that experience a back action from the matter degrees of freedom \cite{Buchler2005, Weimer2010, Keilmann2011, Edmonds2013, Greschner2014, Bermudez2015, Barbiero2019, Cuadra2020}. A principal motivation behind such developments concerns the elucidation of non-perturbative effects in lattice gauge theories (LGTs)~\cite{Kogut1979, Wiese2013, Zohar2015, Dalmonte2016}. First realizations of density-dependent gauge fields in cold gases (which did not satisfy the constraints of a gauge theory), were reported in Refs.~\cite{Chin2018, Gorg2019}. Besides, major advances in the quantum simulation of LGTs have been achieved in trapped ions~\cite{Martinez2016}, and more recently, in ultracold atoms~\cite{Schweizer2019, Mil2020}. 


It is well established that artificial gauge fields can also be engineered in the solid state, for instance, by applying strain to materials~\cite{aidelsburger2018artificial}. In the context of graphene~\cite{CastroNeto2009, Vozmediano2010, Goerbig2011}, strain generates an effective ``magnetic" field, which strongly modifies its low-energy relativistic excitations:~strain induces relativistic Landau levels in the vicinity of the Dirac points~\cite{Guinea2010}. The main and crucial difference with the action of a real magnetic field is that time-reversal symmetry is preserved in strained graphene. As a result, the vector potential that emerges from the strain field has opposite signs at the two valleys, thus providing the conditions for the valley Hall effect~\cite{Xiao2007}. The characteristic relativistic Landau spectrum has been successfully observed in graphene~\cite{Levy2010} and in molecular graphene~\cite{Gomes2012}. In synthetic systems, lattice patterning is often an intrinsic requirement, such that an external stretching is not needed to produce the effects of strain. Instead, strain can be mimicked by displacing the lattice sites according to the most convenient profile 
\cite{Schomerus2013, Salerno2015, Brendel2017, Abbaszadeh2017, Yang2017, Guglielmon2021}. The physics of strained honeycomb lattices has been investigated in photonic systems with arrays of optical waveguides~\cite{Rechtsman2013a}, microwave resonators~\cite{Bellec2020}, exciton-polaritons~\cite{Jamadi2020}, and in acoustic metamaterials~\cite{Wen2019}. In contrast, optical-lattice potentials for ultracold atoms are typically rigid:~their perfect periodicity is generally fixed by the lasers wavelength. This makes the realization of strain more challenging in cold atoms than for other synthetic-matter platforms. We note that a promising proposal, which consists in displacing one of the three laser beams generating the honeycomb-lattice potential, was described in Refs.~\cite{Alba2013, Pekker2015}.


In this paper, we introduce a radically different strategy to realize and investigate the effects of strain in optical lattices, which is summarized in Fig.~\ref{fig:Fig_1}a. Our scheme builds on a mixture of two atomic species, one of which is bosonic (denoted by $\ua$) and forms a Bose-Einstein condensate (BEC), while the second species (denoted by $\da$) can be either bosonic or fermionic. As a central ingredient, the two species are assumed to be coupled through a density-assisted tunneling term, which affects the hopping of $\da$ atoms through the density of $\ua$ atoms. When the BEC is harmonically trapped, the density of $\ua$ atoms is inhomogeneous, and the correlated tunneling of $\da$ atoms displays the effects of a fictitious uniaxial strain:~the $\da$ atoms behave as electrons moving in a strained lattice. We discuss the validity of this scheme for two different regimes of the condensate, namely the non-interacting and the Thomas-Fermi regimes. In both cases, we show that the spectrum associated with $\da$ atoms can display pseudo-Landau levels under proper conditions. We propose a Floquet driving protocol to engineer the required coupling between the two species and we outline probing methods to extract the spectral and topological features. Differently from previously suggested schemes, the achieved strain field is dynamical, and thus, it is closer to the actual solid-state setting where phononic vibrations are present. We note that similar models have been suggested in different contexts, for instance, to design an atomic dissipative bath~\cite{Griessner2007, Bruderer2007}, in order to simulate the Su-Schrieffer-Heeger instability~\cite{Cuadra2018}, to study the back-action of dipolar crystal~\cite{Pupillo2008} and vortex lattice fluctuations~\cite{Chaviguri2017, Chaviguri2018}. Within our framework, fluctuations of the strain field are carried by the density modes of the condensate. This fact opens a route to novel interesting scenarios, in which the back action of the matter degrees of freedom onto the synthetic strain field can be theoretically and experimentally investigated. 

The paper is organized as follows: Section~\ref{sec:strain_honey} reviews the main features of uniaxial linear strain in graphene-type lattices; this introductory material aims at describing the key concepts and phenomena associated with strain-induced magnetic fields, which will be extensively used in the core of our work; in Sec.~\ref{sec:model}, we introduce and solve the model for simulating strain in optical lattices via coupling two atomic mixtures with density-dependent hopping terms; in Sec.~\ref{sec:validation}, we analyze the results by computing the fidelity of the resulting eigenstates; in Sec.~\ref{sec:realization}, we propose a Floquet scheme to implement the model and we discuss several experimentally relevant aspects and probing methods; in Sec.~\ref{sec:conclusions}, we summarize our work and draw our conclusions.


\section{Strain on the honeycomb lattice}\label{sec:strain_honey}

The Hamiltonian of a single-particle on the honeycomb lattice in the tight-binding approximation reads 
\be
\hat H_{0} = - \sum_{\mathbf r \in \mathcal A,j} t_j(\hat a_{\mathbf r}^\dagger \hat b^{}_{\mathbf r + \boldsymbol \delta_j} + \text{h.c.}), \qquad j \in \{1,2,3\},
\ee
where $\hat a_{\mathbf r}^{},\hat a^\dagger_{\mathbf r}$ ($\hat b^{}_{\mathbf r},\hat b^\dagger_{\mathbf r}$) are respectively the annihilation and creation operators at position $\mathbf r \equiv (x,y)$ in the $\mathcal A$ ($\mathcal B$) sublattice, the quantities $t_j$ are the nearest-neighbor hopping amplitudes and $\boldsymbol\delta_1\!=\!(-a,0)$, $\boldsymbol\delta_2\!=\!(a/2,\sqrt 3 a/2)$, $\boldsymbol\delta_3 \!=\! (a/2,-\sqrt 3 a/2)$, as in Fig.~\ref{fig:Fig_1}a. In momentum space, $\hat H_0$ can be rewritten as \begin{equation}\label{key}
\hat H_0 = \sum_{\mathbf k} (\hat a_\mathbf{k}^\dagger, \hat b^{\dagger}_\mathbf{k})\, h(\mathbf k) \begin{pmatrix}
\hat a_\mathbf{k}\\
\hat b_\mathbf{k}
\end{pmatrix},
\end{equation} 
with
\be
\label{eq:hamk}
h(\b k) = 
\begin{pmatrix}
	0 & -\sum_j t_j e^{i \mathbf k \cdot \boldsymbol \delta_j}\\ 
	-\sum_j t_j e^{-i \mathbf k \cdot \boldsymbol \delta_j} &0 
\end{pmatrix}.
\ee
Assuming $C_3$ discrete rotational invariance, namely $t_l = t$, the Hamiltonian around the time-reversal invariant points $\mathbf K$ and $-\mathbf K$ in the Brillouin zone shown in Fig.~\ref{fig:Fig_1}b reads
\begin{equation}\label{eq:Dirac}
h(\mathbf q, \zeta \mathbf K) = \hbar v_\text{F} (\zeta q_y \sigma_x - q_x \sigma_y),
\end{equation}
where $\zeta = \pm 1$, $\mathbf q \equiv \mathbf k-\zeta \mathbf K$, $v^{}_\text{F}\equiv 3ta/2\hbar$ is the Fermi velocity and $\sigma_x,\sigma_y$ are Pauli matrices. Eq.~\eqref{eq:Dirac} describes a relativistic Dirac particle whose linear dispersion relation is given by
\be
\begin{split}
	\epsilon(\mathbf q) = \pm \hbar v_\text{F} |\mathbf q|.
\end{split}
\ee
The application of a spatial deformation that changes the distance between lattice sites, also known as strain, brings new interesting effects \cite{Vozmediano2010,CastroNeto2009,Salerno2015}. Within the tight-binding description and for small deformations, strain affects the tunneling amplitudes, which become spatially dependent as $t_j \rightarrow t_j(\b r)$. Different types of strain can be applied to the honeycomb lattice \cite{Salerno2017}, but here we will consider the case of uniaxial linear strain along the $x$ direction. For an intensity of strain $\tau \ll 1$, we assume that the hopping coefficients read
\be\label{t_j}
\begin{split}
	t_j(x) = t\left(1+\tau \frac{(x-x_c)}{3a^2} |\hat{\mathbf x} \cdot \boldsymbol{\delta}_j |\right), \quad \hat{\mathbf x} = (1,0),
\end{split}
\ee
with the condition $\tau L_x/3a < 1$ ensuring that the strain is sufficiently small to avoid a local Lifshitz transition to a gapped state \cite{Salerno2015}. By introducing this slow space dependence of the hopping coefficients into the Hamiltonian \eqref{eq:hamk}, translational invariance along $y$ is preserved and $k_y$ remains a good quantum number. In the rest of this work, we will exploit translational invariance by solving $\hat H_0$ for a stripe of size $N_x \times 2$, as shown in Fig.~\ref{fig:Fig_1}a where the unit cell of the $y$-periodic lattice is highlighted.

\begin{figure}[!t]
	\center
	\includegraphics[width=0.9\columnwidth]{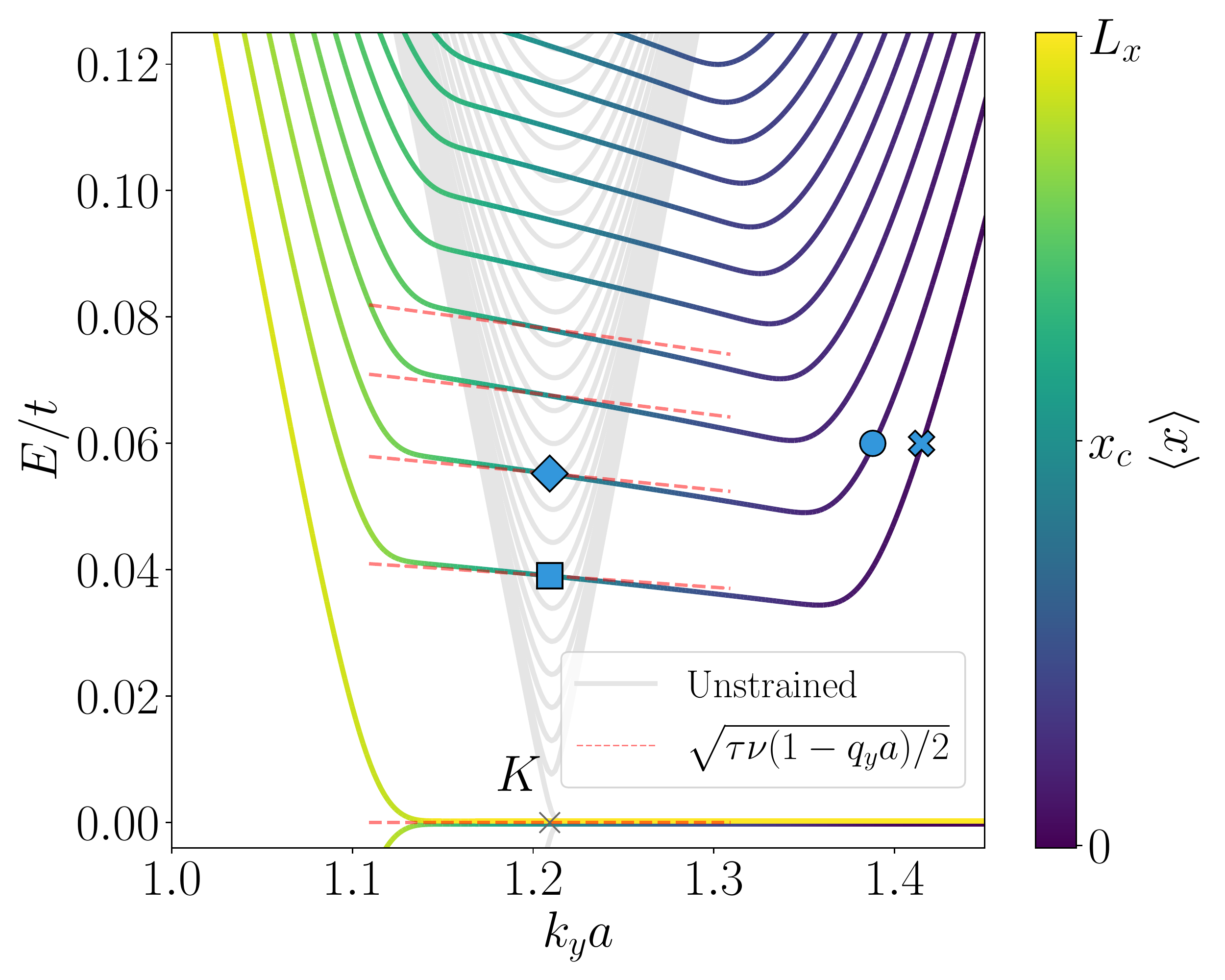}
	\caption{Spectrum of $\hat H_0$ for $N_x = 601$ sites, $\tau = 0$ (in light gray) and $\tau = 0.003$ (gradient of colors). In the strained case, the color represents the mean position of each eigenstate. The four dots with different markers correspond to the eigenvectors plotted in Fig.~\ref{fig:Fig_3} using the corresponding marker.}
	\label{fig:Fig_2}
\end{figure}

\begin{figure}[!t]
	\center
	\includegraphics[width=1.0\columnwidth]{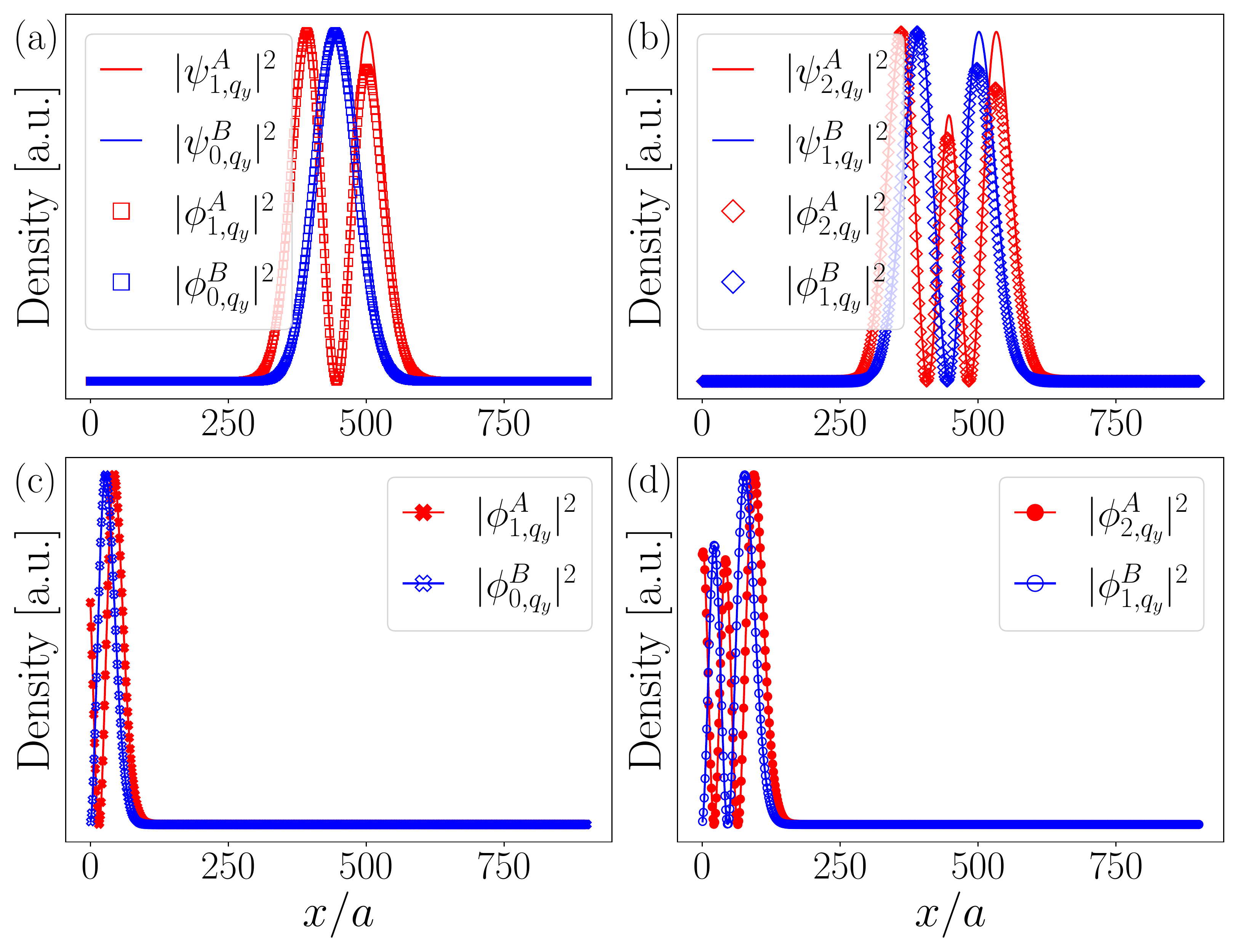}
	\caption{Density corresponding to the eigenfunctions of  $\hat H_{0}$ for $N_x = 601$, $\tau = 0.003$ for various values
	 of $(q_y a,E)$: (a) $(0,0.039)$, (b) $(0,0.055)$, (c) $(0.206,0.06)$, (d) $(0.179,0.06)$. Their respective markers correspond to the ones in Fig.~\ref{fig:Fig_2}. While panels (a) and (b) show bulk eigenfunctions of the first and second Landau levels, panel (c) and (d) show corresponding edge states localized at the left edge near $x=0$. 
	 }
	\label{fig:Fig_3}
\end{figure}

Uniaxial linear strain on the honeycomb lattice mathematically appears in the Dirac Hamiltonian~\eqref{eq:Dirac} as a homogeneous magnetic field \cite{Vozmediano2010}
\begin{equation}\label{H_Dirac}
h(\mathbf q,\zeta \mathbf K,\mathbf A) = \hbar v_\text{F} ((\zeta q_y-e^*A_y) \sigma_x - (q_x - e^*A_x)\sigma_y)\,,
\end{equation}
where $\mathbf A$ has the form of a vector potential in the Landau gauge
\begin{equation}\label{A_Dirac}
\begin{split}
e^*\mathbf A &= (0,\zeta(2t_1-t_2-t_3)/2v_\text{F})\\
&= \left(0,\zeta \frac{\hbar \tau }{9a^2} (x-x_c)\right).
\end{split}
\end{equation}
In the rest of this work, we will use units where $e^* = 1$. Since time-reversal symmetry is preserved by strain, the corresponding magnetic field $\mathbf B = \boldsymbol \nabla \times \mathbf A$ has opposite  sign for the two valleys \cite{Vozmediano2010,Goerbig2011}. As in the non-relativistic case, the spectrum of a relativistic particle in a magnetic field also displays Landau levels (LLs). A major difference with respect to their nonrelativistic counterparts is that they are not equispaced in energy. The full expression, which includes a momentum dependence originating from a spatially varying Fermi velocity \cite{Salerno2015}, reads
\be\label{tiltedLL}
\epsilon^{\text{LL}}_{\nu} (q_y) = \pm t\sqrt{\nu \frac{\tau}{2}} \sqrt{1-\zeta q_y a},  \quad \nu \in \mathbb N,
\ee
corresponding to the relativistic LL eigenvectors $\Psi_{\nu,q_y} = (\psi_{\nu,q_y}^A,\psi_{\nu,q_y}^B)$, where
\begin{equation}\label{Landaustates}
\psi^l_{\nu,q_y}(\mathbf r) \propto e^{iq_y y}e^{-\frac{(x-x_0)^2}{2\ell_B^2}} H_{\nu}\left(\frac{x-x_0}{\ell_B}\right),
\end{equation}
with $x_0(q_y) \equiv x_c-\zeta q_y \ell_B^2$ indicating the LL center, $x_c$ the origin of the coordinate axis and $l = A,B$ the component of the wavefunction associated to the $\mathcal A$ or $\mathcal B$ sublattice, respectively. The function $H_{\nu}$ is the $\nu^\text{th}$ Hermite function and $\ell_B$ is the magnetic length, related to the strain parameter by
\begin{equation}\label{key}
\ell_B = \frac{3a}{\sqrt \tau}.
\end{equation}
In Fig.~\ref{fig:Fig_2}, we compare the numerically calculated spectrum of $\hat H_0$ in the presence of uniaxial linear strain with the one in the absence of strain, near the $K$ point.  We observe that straining the lattice has generated relativistic LLs as predicted by Eq.~\eqref{tiltedLL}.  In Figs.~\ref{fig:Fig_3}a,b, we compare the eigenstates of $\hat H_0$ corresponding to the first and second LLs at $k_ya = 2\pi/3\sqrt 3$ (or $q_y = 0$), denoted $\phi_{\nu,q_y} = (\phi^A_{\nu,q_y},\phi^B_{\nu-1,q_y})$, with the analytical relativistic Landau states $\Psi_{\nu,q_y}$, for $\nu=1$ and $\nu=2$ respectively. Sufficiently far from $K$, the (almost) flat LLs become strongly dispersive. This effect originates from the dependence of the LL wavefunction center on momentum, $x_0(q_y)$, which we show in Figs. \ref{fig:Fig_3}c,d. For the values of $q_y$ corresponding to wavefunctions centered near the edge of the system, the hardwall potential lifts these states in energy thus causing a strong dispersion. The energy levels therefore cross the energy gap between two subsequent Landau levels. This is indeed what we expect from the bulk-boundary correspondence for the quantum Hall effect (QHE) that predicts the existence of robust edge modes when the Fermi level sits in the gap between two LLs. Since the dispersion shows opposite slope at the two opposite edges, these modes are obviously chiral. However, we have to recall that on the other valley, the opposite effect will take place as a result of time-reversal symmetry \cite{Salerno2017}. In the end, no net current can be observed on each edge, unless valley transport can be resolved, an effect known as valley Hall effect.

\section{Model}\label{Model}
\label{sec:model}
In order to generate uniaxial linear strain with cold atoms in optical lattices, we propose to employ a mixture of two atomic species, which we indicate as $\uparrow$ and $\downarrow$. The $\uparrow$ atoms are weakly interacting bosons, harmonically trapped in the $x$ direction. The corresponding Hamiltonian reads
\be\label{H_B}
\begin{split}
	&\hat H_{\uparrow} = -J \sum_{\mathbf{r} \in \mathcal A,j} \left( \hat{a}^\dagger_{\uparrow,\mathbf{r}} \hat b_{\uparrow,\mathbf r + \boldsymbol \delta_j}^{} + \text{h.c.} \right)\\
	&+ \frac{V_x}{2} \sum_{\mathbf r\in \mathcal A,\mathcal B} (x-x_c)^2 \hat n_{\uparrow,\mathbf r}+ \frac U2 \sum_{\mathbf r \in \mathcal A,\mathcal B} \hat n_{\uparrow,\mathbf r} (\hat n_{\uparrow,\mathbf r} - 1),
\end{split}
\ee
where $\hat n_{\uparrow,\mathbf r} \equiv \hat a^\dagger_{\uparrow,\mathbf r} \hat a^{}_{\uparrow,\mathbf r}$ ($\hat b^\dagger_{\uparrow,\mathbf r}\hat b^{}_{\uparrow,\mathbf r}$) if $\mathbf r \in \mathcal A$ ($\in \mathcal B$) and $x_c$ is the position of the system's center. The parameters $J$, $U$ and $V_x$ are respectively the nearest-neighbor hopping amplitude, the onsite interaction energy and the strength of the harmonic confinement. 

The $\downarrow$ atoms, whose statistics does not need to be specified, hop on the same honeycomb lattice as the $\uparrow$ atoms according to the Hamiltonian
\begin{equation}\label{key}
\begin{split}
\hat H_\downarrow &=  -t\sum_{\mathbf r \in \mathcal A,j} \hat a^\dagger_{\downarrow,\mathbf{r}} \hat b^{}_{\downarrow,\mathbf r + \boldsymbol \delta_j} + \text{h.c.},
\end{split}
\end{equation}
where $t$ is the hopping amplitude for the $\downarrow$ atoms. The two species are coupled through the interaction term
\begin{equation}\label{key}
\begin{split}
\hat H_{\uparrow \downarrow} &= -\alpha t \sum_{\mathbf r \in \mathcal A ,j} \hat a^\dagger_{\downarrow,\mathbf{r}} F_j(\hat n_{\uparrow,\mathbf{r}},\hat n_{\uparrow,\mathbf r + \boldsymbol \delta_j}) \hat b^{}_{\downarrow,\mathbf r + \boldsymbol \delta_j} + \text{h.c.},
\end{split}
\end{equation}
where $\alpha$ is a dimensionless parameter quantifying the interaction strength between the $\uparrow$ and $\downarrow$ species. Since the functions $F_j$ depend on the density of the bosons, this term describes \textit{correlated hopping} (or \textit{density-assisted}) processes where the tunneling of $\downarrow$ atoms between two neighboring sites depends on the number of $\uparrow$ atoms at these two sites. For the functions $F_j$, we consider the following expression
\be\label{f_n}
\begin{split}
	 F_j(\hat n_{\uparrow,\mathbf{r}},\hat n_{\uparrow,\mathbf{r}+\bs\delta_j}) = \frac{1}{3} \sigma_j (\hat n_{\uparrow,\mathbf r + \boldsymbol \delta_j}-\hat n_{\uparrow,\mathbf{r}}),
\end{split}
\ee
where $\sigma_1 = 1$ and $\sigma_2 = \sigma_3 = -1$. As we show below, these functions will generate the artificial strain for the $\downarrow$ atoms when the density of $\uparrow$ atoms is inhomogeneous. The full model reads\begin{equation}\label{key}
\hat H = \hat H_\uparrow +\hat H_{\downarrow} + \hat H_{\uparrow \downarrow},
\end{equation}
which we solve in the mean-field (MF) approximation for the $\uparrow$ atoms, described by the discrete Gross-Pitaevskii (GP) equation, and by neglecting the back action of the $\downarrow$ atoms on the condensate. Within this approximation, the BEC of $\uparrow$ atoms acts as a background for the $\downarrow$ atoms and we therefore write the model as $\hat H \simeq \hat H^{\text{eff}}_{\downarrow} + \hat H^{\text{MF}}_{\uparrow}$,  where $\hat H^{\text{eff}}_{\downarrow} \equiv \hat H_{\downarrow} + \hat H^{\text{MF}}_{\uparrow\downarrow}$. The density operator $\hat n_{\uparrow,\mathbf r} $ is then replaced by its mean value $\bar n_{\uparrow}(x)$, where we remove the dependence on $y$ due to the assumption of homogeneity in this direction. After calculating the BEC density profile obtained by solving $\hat H^{\text{MF}}_\uparrow$, we input the solution into $\hat H_\downarrow^{\text{eff}}$. As a result, the $\downarrow$ atoms experience spatially dependent hopping parameters $t^{\text{eff}}_j$ that read
\begin{equation}\label{effective_tj}
\begin{split}
t^{\text{eff}}_1(x) &=t\left[1-\frac{\alpha}{3} (\bar n_{\uparrow}(x)-\bar n_{\uparrow}(x-a)) \right],\\
t^{\text{eff}}_{2,3}(x) &=t\left[1-\frac{\alpha}{3} \left(\bar n_{\uparrow}\left(x+ a/2\right)-\bar n_{\uparrow}(x)\right)\right].
\end{split}
\end{equation}
In order to recover the linear space dependence needed for uniaxial strain, we consider a parabolic profile
\begin{equation}\label{parab_profile}
\bar n_\uparrow(x) = -\eta_1 \frac{(x-x_c)^2}{a^2} + \eta_0\,,
\end{equation}
where the constants $\eta_0$, $\eta_1$ will be specified below and depend on the microscopic parameters of the BEC regime considered. As a result, we obtain
\begin{equation}\label{t_jeta1}
\begin{split}
t_1^{\text{eff}} &= t \left[1+ \frac 23 \frac{\eta_1 \alpha}{a} (x-x_c) - \frac{\alpha \eta_1}{3} \right],\\
t_{2,3}^{\text{eff}} &= t \left[1+ \frac{1}{3} \frac{\eta_1 \alpha}{a} (x-x_c) + \frac{\alpha \eta_1 }{12} \right],
\end{split}
\end{equation}
which reproduces uniaxial linear strain with $\tau = 2\alpha \eta_1$, as described by Eq.~\eqref{t_j}. We obtain new constant terms appearing in Eq.~\eqref{t_jeta1} in comparison to Eq.~\eqref{t_j}, which result in the vector potential
\begin{equation}\label{A_corr}
\mathbf A = \left(0,\frac{2\zeta\hbar\alpha\eta_1 }{9a^2}(x-x_c) -\frac{5\zeta \hbar \alpha\eta_1}{36a}\right).
\end{equation}
The last term in Eq.~\eqref{A_corr} shifts the position of the Dirac points. However, such a shift is negligible as long as $\alpha \eta_1 \ll 1$, which is the case here. The effect of this term is indeed not observable in the numerical results presented below. Furthermore, note that the LLs energy gaps are not affected because the magnetic field is determined by the derivatives $\partial_x t_j^\text{eff}$. 

\begin{figure}[!t]
	\center
	\includegraphics[width=0.7\columnwidth]{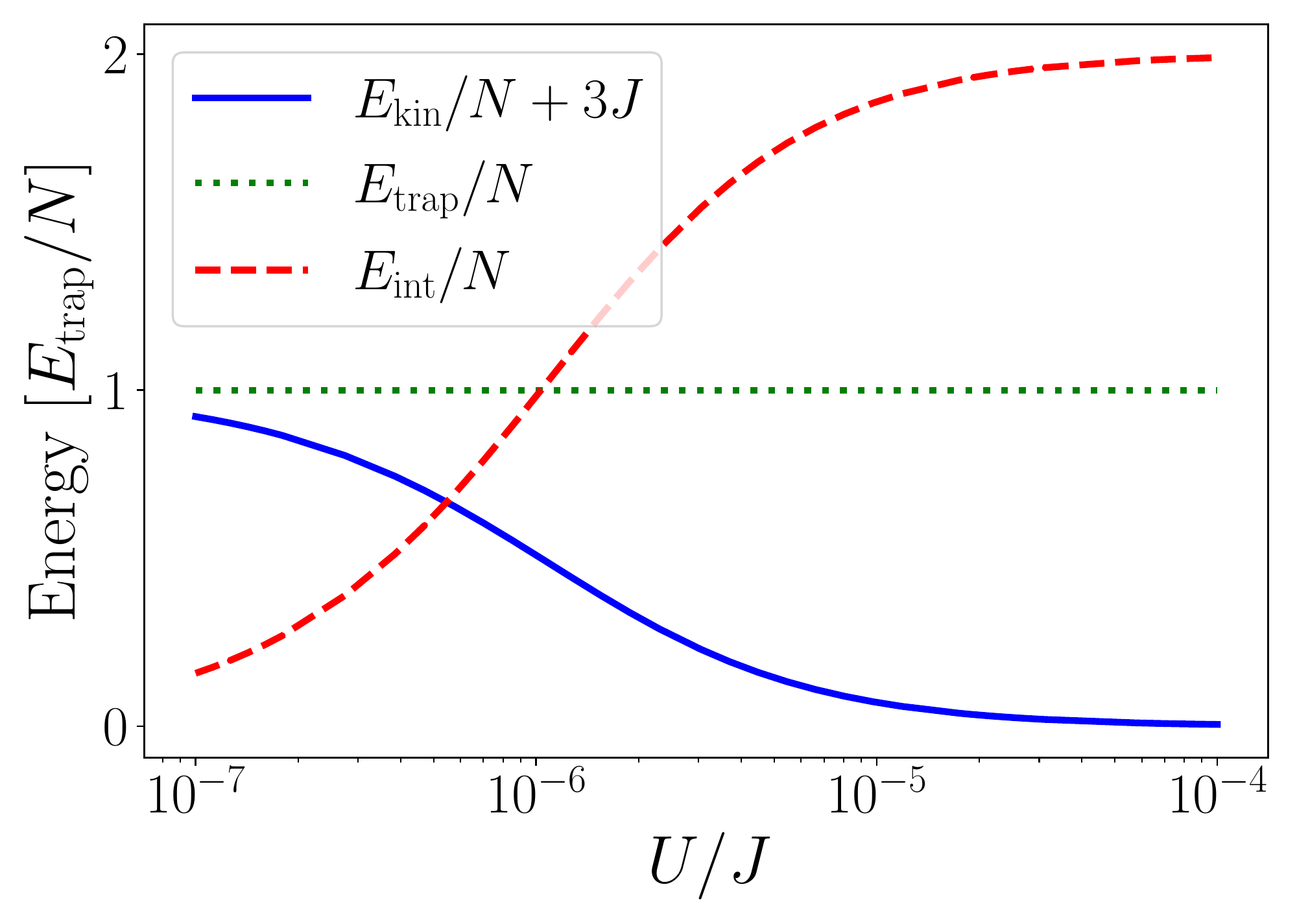}
	\caption{Comparison of the energy contributions (in units of the harmonic trap energy) for $N_x = 601$,  $V_x = 10^{-6}J/a^2$ and $N^\ua = 1.2 \times 10^5$. The TF regime is reached when the kinetic energy is negligible with respect to the other two contributions, i.e. for sufficiently high values of $U$. The kinetic energy is shifted by $3J$ in order to measure the energy from the bottom of the band.}
	\label{fig:Fig_4}
\end{figure}

In the rest of this work, we discuss two regimes where the density profile can be approximated by a parabolic expression Eq.~\eqref{parab_profile}: the non-interacting and the Thomas-Fermi (TF) regimes. The former corresponds to the condition $U = 0$, whereas the latter is obtained for large $U$ such that the BEC kinetic energy becomes negligible. In order to correctly identify the TF regime, we compare the energy functionals
\begin{equation}\label{E_contr}
\begin{split}
E_\text{trap}[\Psi_\uparrow] =& \frac{V_x }{2}\sum_{\mathbf r \in \mathcal{A,B}} \left(x-x_c\right)^2\,|\Psi_{\uparrow,\mathbf r}|^2,\\
E_\text{kin}[\Psi_\uparrow] =& -J \sum_{\mathbf r \in \mathcal{A,B},j} \left(\Psi^*_{\uparrow,\mathbf r}  \Psi_{\uparrow,\mathbf r+\boldsymbol{\delta}_j} + \text{h.c.} \right),\\
E_\text{int}[\Psi_\uparrow] =& \frac{ U}{2} \sum_{\mathbf r \in \mathcal{A,B}} \ |\Psi_{\uparrow,\mathbf r}|^2(|\Psi_{\uparrow,\mathbf r}|^2-1) .
\end{split}
\end{equation}
with each other, where $\Psi_{\uparrow,\mathbf r} = \braket{\hat a_{\uparrow,\mathbf r}}$ for $\mathbf r \in \mathcal A$ and $\Psi_{\uparrow,\mathbf r} = \braket{\hat b_{\uparrow,\mathbf r}}$ for $\mathbf r \in \mathcal B$. The system enters the TF regime when $E_\text{kin}\ll E_\text{trap},E_\text{int}$ \cite{Pitaevskii2016}, which is reached for sufficiently large values of $U$, as shown in Fig.~\ref{fig:Fig_4}.

\begin{figure}[!t]
	\includegraphics[width=0.5\textwidth]{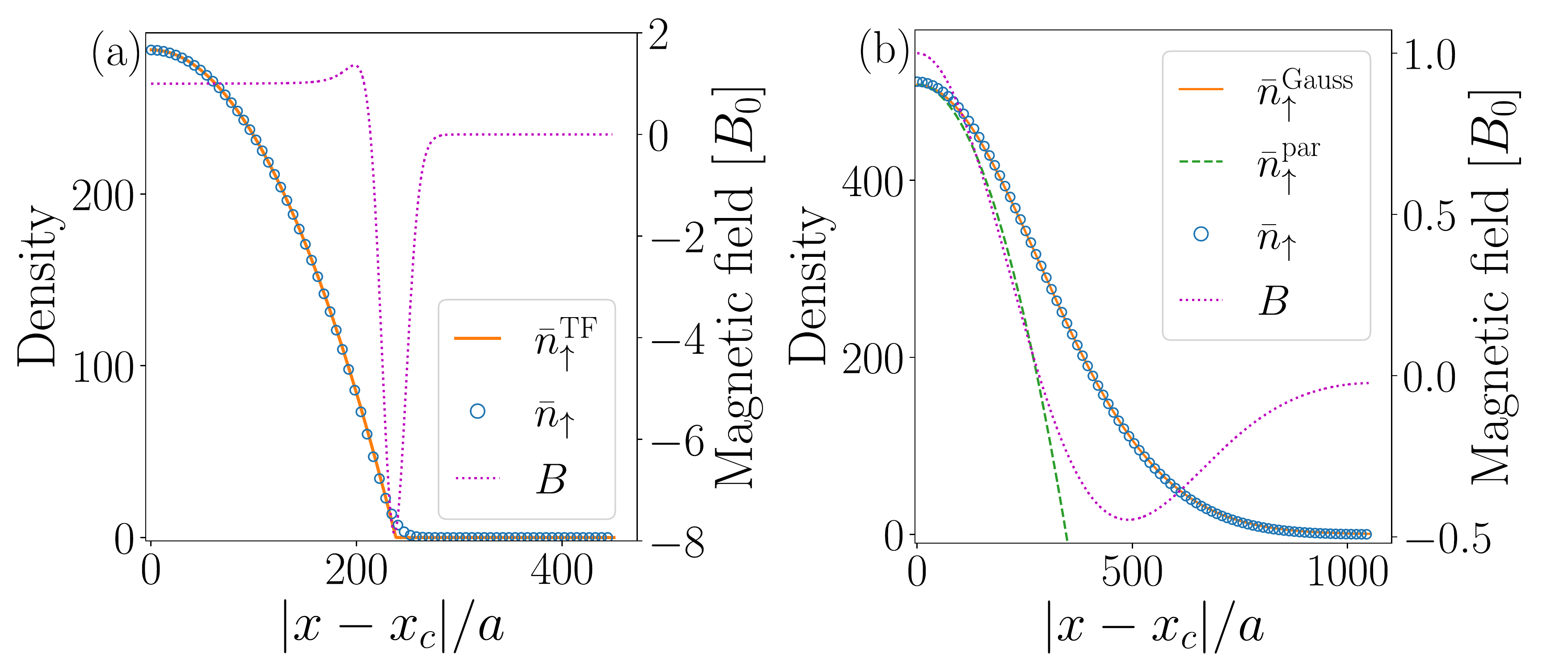}
	\caption{(a) Density profile of the $\ua$ atoms in the TF regime for $V_x = 10^{-6}J/a^2$, $N^\ua = 1.2 \times 10^5$, $N_x = 601$ and $U = 10^{-4} J$. (b) Density profile of the $\ua$ atoms in the non-interacting regime for $V_x = 5.8 \times 10^{-11} J/a^2$, $N^\ua = 4.8 \times 10^5$, $N_x = 1401$. The dashed line is the approximate parabolic profile valid for $|x-x_c| \ll \xi$, where $\xi$ is the Gaussian width (see Eq.~\eqref{parab_U0}). The resulting magnetic field (dotted line) is shown in both panels in units of $B_0$, the value of the magnetic field at the center of trap, $x_c$.
	}
	\label{fig:Fig_5}
\end{figure}

These two regimes are going to be the focus of our analysis, as they will lead to inhomogeneous hopping coefficients for the $\downarrow$ atoms described by Eq.~(\ref{effective_tj}). The BEC density profiles are shown in Fig.~\ref{fig:Fig_5}, where we also show the corresponding magnetic fields obtained from Eq.~\eqref{A_Dirac},
\begin{equation}\label{key}
B(x) = \frac{\zeta}{2v_\text{F}} \partial_x (2t_1^\text{eff} - t_2^\text{eff} -t_3^\text{eff})\,,
\end{equation}
calculated by neglecting the space dependence of the Fermi velocity. 

\subsection{Thomas-Fermi regime for the $\uparrow$ atoms}

\begin{figure}[!b]
	\includegraphics[width=0.9\columnwidth]{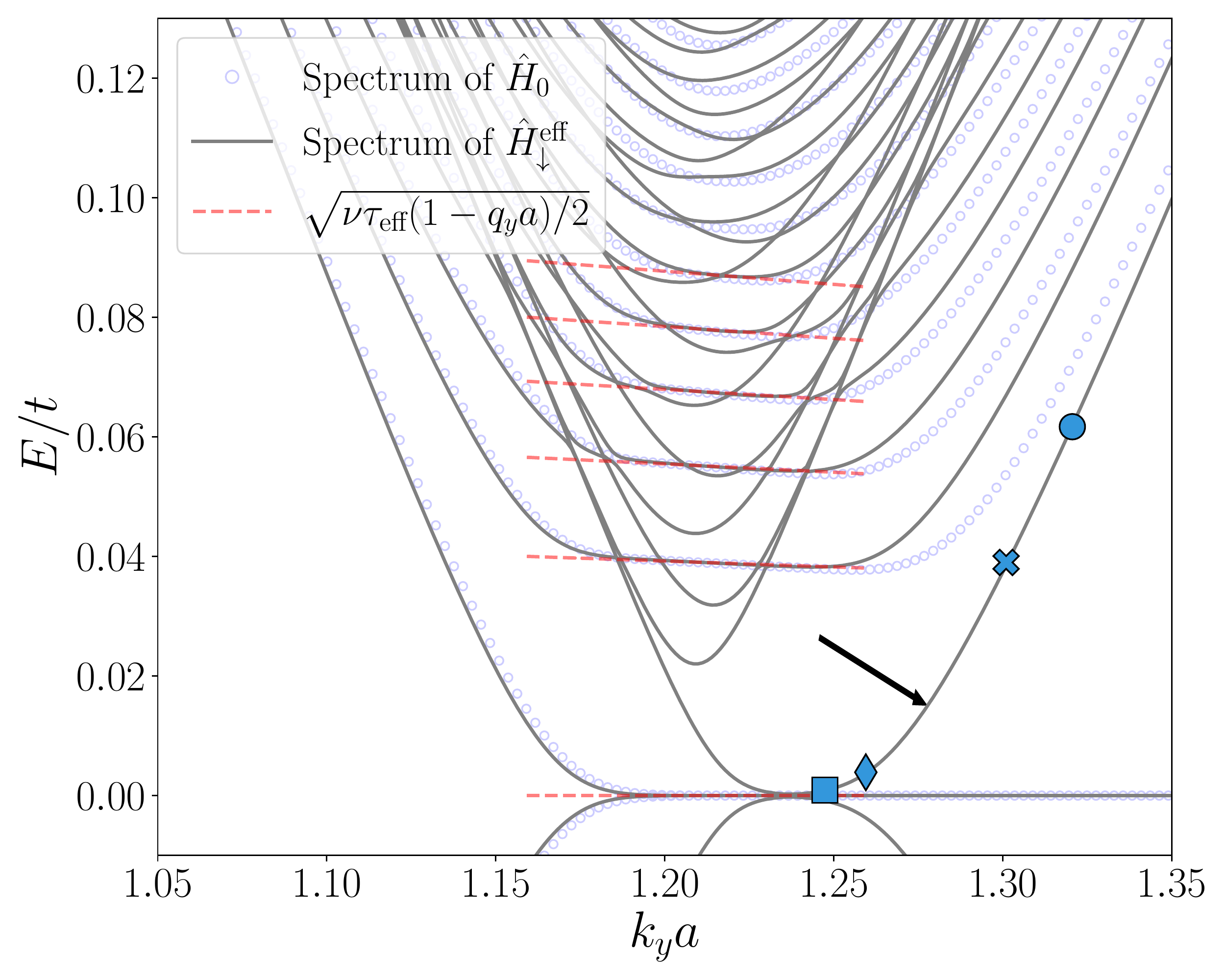}
	\caption{Spectrum of $\hat H^\text{eff}_\downarrow$ when the BEC of $\ua$ atoms is in the TF regime, including a comparison with the spectrum of the strained graphene Hamiltonian $\hat H_0$ (empty circles) and the LLs predicted by Eq.~\eqref{tiltedLL} (dashed lines). The number of sites along $x$ is $N_x = 601$ and $U = 10^{-4}J$, $V_x = 10^{-6}J/a^2$, $N^\ua = 1.2 \times 10^5$, $\alpha = 0.3$. The strain intensity is $\tau_{\text{eff}} = 0.003$. The four dots with different markers correspond to the eigenvectors plotted in Fig.~\ref{fig:Fig_7} using the corresponding marker.}
	\label{fig:Fig_6}
\end{figure}

We start our analysis by investigating the TF regime, which the gas of $\uparrow$ atoms enters when the repulsive interactions dominate the kinetic energy. By inspecting Fig.~\ref{fig:Fig_4}, which is obtained for $V_x =  10^{-6} J/a^2$ and a number of atoms per stripe  $N^\ua = 1.2 \times 10^5$, we see that we can safely use the TF approximation for $U \gtrsim 10^{-5}J$. The corresponding density profile reads
\begin{equation}\label{n_TF}
\bar n_{\uparrow}(x) = \frac{1}{U} \left[\mu - \frac{V_x}{2} (x-x_c)^2\right],
\end{equation}
where $\mu$ is the BEC chemical potential, which we numerically compute from the relation $\mu = E_{\text{tot}}[N+1]-E_\text{tot}[N]$, where $E_\text{tot}=E_\text{kin}+E_\text{trap}+E_\text{int}$ is the total energy of the BEC. By substituting Eq.~\eqref{n_TF} into Eq.~\eqref{effective_tj}, we find that the effective strain intensity can be expressed in terms of the harmonic trap parameter $V_x$ and the interaction strength $U$ as follows,
\begin{equation}\label{tau_int}
\tau_{\text{eff}} =  \frac{\alpha V_x a^2}{U}\,.\end{equation}
In Fig.~\ref{fig:Fig_6}, we show the spectrum of the $\downarrow$ atoms for $U = 10^{-4} J$. The agreement between the numerical results and the analytical predictions in  Eq.~\eqref{tiltedLL} obtained for $\tau = \tau^\text{eff}$ is visible for the first five levels. 

By looking at the BEC density in Fig.~\ref{fig:Fig_5}a, we see that the TF radius $R_\text{TF} \equiv \sqrt{2\mu/V_x}$ marks a separation between a region with strain ($|x-x_c|<R_\text{TF}$) and a region without strain ($|x-x_c|>R_\text{TF}$). As a consequence, the spectrum displayed in Fig.~\ref{fig:Fig_6} will also show features of a homogeneous (\emph{i.e.} unstrained) honeycomb lattice, as we can conclude by comparison with Fig.~\ref{fig:Fig_2} (gray lines). The interface between strained and unstrained regions is not a hard wall potential, thus allowing for a penetration length of the wavefunctions from both sides. This fact will therefore induce hybridization events between LLs and planewave states that will result in avoided crossings, some of which are visible in Fig.~\ref{fig:Fig_6}. To distinguish the contribution of the two regions in Fig.~\ref{fig:Fig_6}, we superimpose the spectrum (empty circles) of a strained honeycomb system that only extends over the size of the BEC, namely $L_x = 2R_\text{TF}$ with the corresponding value of the strain intensity $\tau = 0.003$. As expected, the LL plateaus are clearly identified together with the edge states branches on the left side of the spectrum ($k_y a <K_y$). Notice that the TF radius sharp boundary has been suggested to host edge modes in other topological interacting models as for the case of spinful bosons, see Ref.~\cite{Galilo2017}. 

 \begin{figure}[!t]
	\center
	\includegraphics[width=\columnwidth]{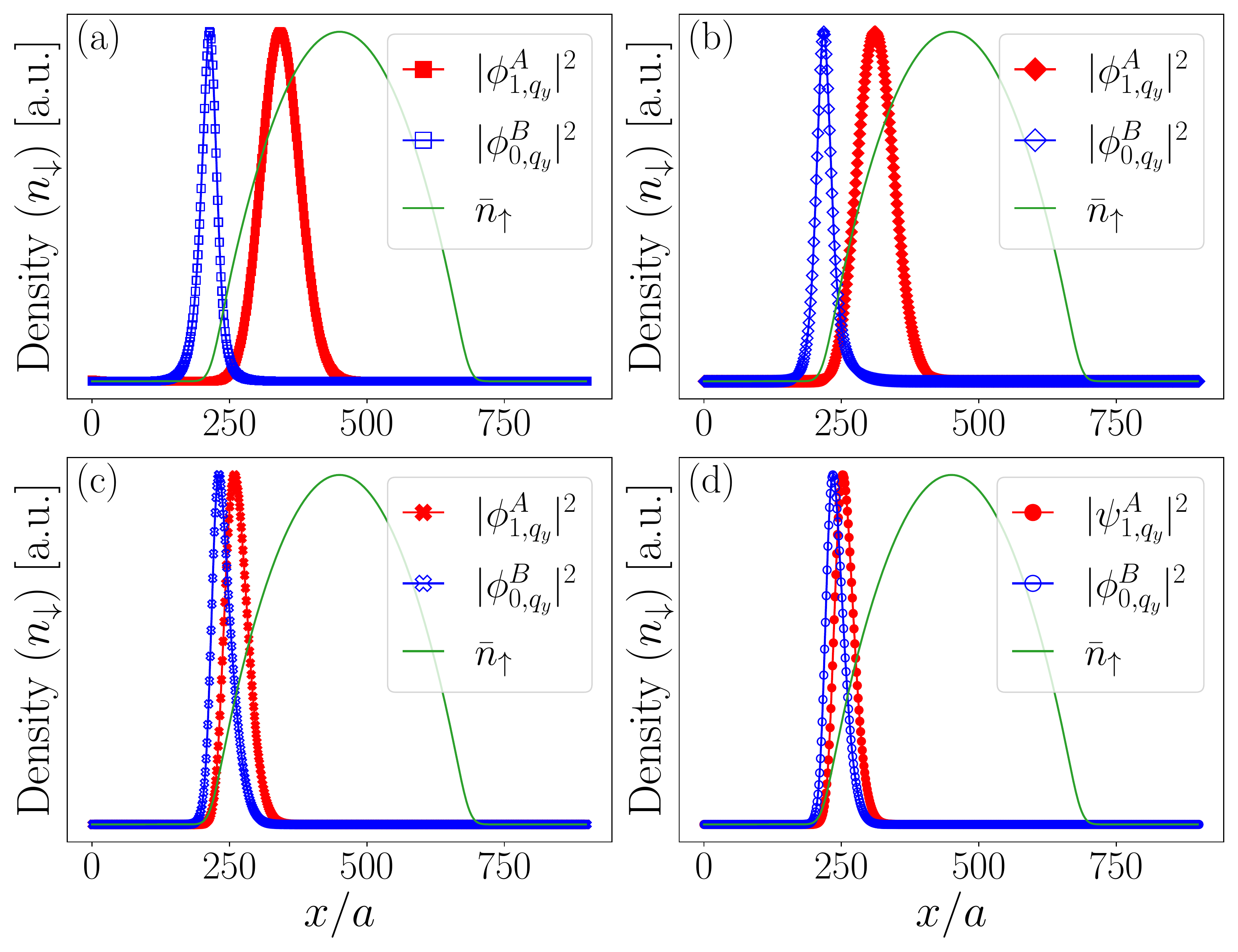}
	\caption{Density of the $\downarrow$ atoms corresponding to the four markers  indicated in Fig.~\ref{fig:Fig_6}. The $A$ and $B$ components of the eigenvectors of $\hat H^\text{eff}_\uparrow$ are denoted by $\psi^l_\uparrow$ for $l=A,B$ respectively. Their respective points in the spectrum have the following coordinates $(q_y a,E)$: (a) $(0.038,0.0009)$, (b) $(0.050,0.0039)$, (c) $(0.091,0.0390)$, (d) $(0.111,0.0617)$.}
	\label{fig:Fig_7}
\end{figure}

Deviations from the ideal strain physics appear on the right side of the spectrum for $k_y a>1.25$ and, as indicated by the arrow in the spectrum, a distinct novel branch is present. As pointed out before, the TF radius introduces a separation, or an interface between the two regions. Let us focus, for simplicity, on the left interface at $x \simeq 230 a$. If a hard wall were present and we could therefore cut the system into two separate parts, we would have a zig-zag termination for the unstrained region, which admits $E=0$ edge states for $k_y > K_y$, and LL edge states at energies above the gap for sufficiently large values of $k_y$.  However, since the interface is soft, a hybridization between these two types of states takes place, which results in \emph{i)} a gap-crossing energy branch, as indicated by the arrow in Fig.~\ref{fig:Fig_6} and \emph{ii)} a deviation from the ideal case of strained honeycomb lattice, as manifested by the empty markers in Fig.~\ref{fig:Fig_6} not overlapping with the lines of the spectrum. We show in Fig.~\ref{fig:Fig_7} that the new branch corresponds to LL edge states in the asymptotic limit of large $k_y$, which is also spectrally observed in Fig.~\ref{fig:Fig_6}. A more quantitative analysis of the interface problem, which is beyond the scope of this work, can be addressed by employing the formalism presented in Ref.~\cite{DellAnna2021}, which would describe the localized solutions for a relativistic particle near the interface between a region with magnetic field and a region without. However, additional care must be taken near the TF radius, as the profile smoothens out, thus causing sharp changes in the magnetic field at the interface, which are visible in Fig.~\ref{fig:Fig_5}a.

\subsection{Non-interacting regime for the $\uparrow$ atoms}

For the non-interacting regime, we take $U=0$ and we solve the single-particle Hamiltonian for the $\ua$ atoms. As we will show below, this regime is less ideal for the LL physics, and this is the reason why we present it after the TF regime analysis. The density profile corresponds to the harmonic oscillator ground state, namely a Gaussian profile, as shown in Fig.~\ref{fig:Fig_5}b, which reads
\be
\bar n_\uparrow(x) = \frac{3N^\ua}{4\sqrt{2\pi}\xi}\exp\left(-\frac{(x-x_c)^2}{2\xi^2}\right),\quad \xi \equiv \left(\frac{3Ja^2}{8V_x}\right)^{1/4},
\ee
where $\xi$ is the width of the cloud, $x_c$ is the trap minimum position, $N^\ua$ is the total number of $\uparrow$ atoms and the factor $3/4$ is a consequence of the honeycomb geometry. Near the center of the system, i.e. for $|x-x_c| \ll \xi$, the density can be approximated by the parabola
\be\label{parab_U0}
\bar n_\uparrow^{\text{par}}(x) = \frac{3N^\ua a}{4\sqrt{2\pi} \xi} \left[1-\frac{(x-x_c)^2}{2\xi^2} \right].
\ee
By substituting Eq.~\eqref{parab_U0} into Eq.~\eqref{effective_tj} for $t_j^{\text{eff}}$, we obtain the strain parameter
\be\label{tau_U0}
\tau_{\text{eff}} = \frac{3 N^\ua\alpha a^3}{4\sqrt{2\pi}\xi^3}. 
\ee

\begin{figure}[!t]
	\center
	\includegraphics[width=0.9\columnwidth]{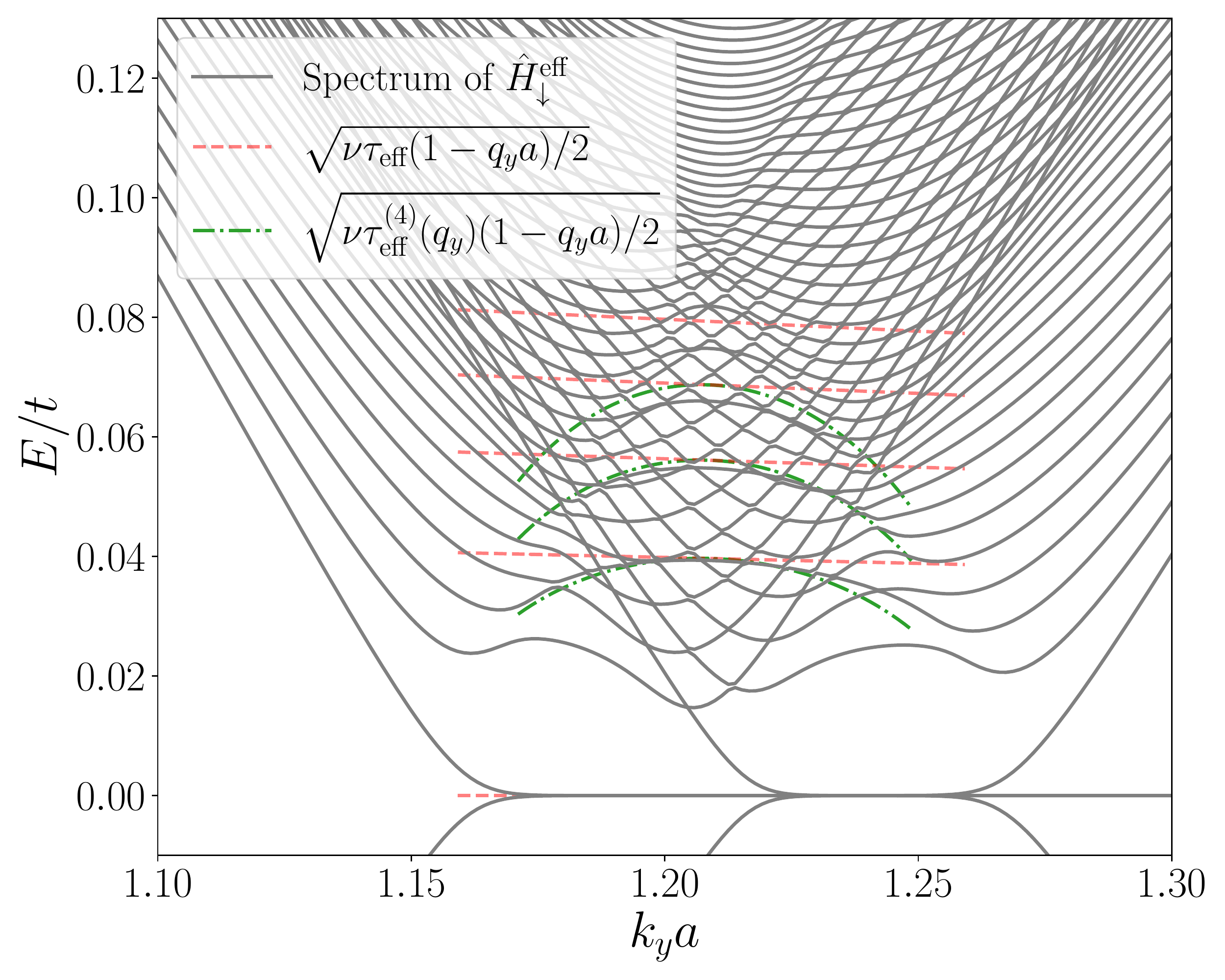}
	\caption{Spectrum of $\hat H_\downarrow^\text{eff}$ when the BEC of $\ua$ is in the non-interacting regime. The number of sites along $x$ is $N_x = 1401$, $V_x = 5.8 \times 10^{-11}J/a^2$, $N^\ua = 4.8 \times 10^5$ and $\alpha = 0.49$. The strain intensity is $\tau_{\text{eff}} = 3.1\times 10^{-3}$. Dashed lines represent the predicted LL energies for a homogeneous synthetic magnetic field, whereas the dashed-dotted lines are obtained by including the leading effects of inhomogeneity.}
	\label{fig:Fig_8}
\end{figure}

Differently from the TF regime, the analytical prediction in Eq.~\eqref{tiltedLL} is in good agreement with the numerical spectrum only very close to the $K$ point, as shown in Fig.~\ref{fig:Fig_8}. Important differences with the TF regime appear in the spectrum, whose origin can be identified directly from the inspection of the two types of density profiles (see Fig.~\ref{fig:Fig_5}) and traced back to \emph{i)} the severe deviations of the noninteracting density profile from an ideal parabola and \emph{ii)} the absence of a sharp transition from a region with strain to a region without it. As a result, we obtain a nonlinear space-dependence of strain, which translates into an inhomogeneous magnetic field that decreases in strength away from the center, see Fig.~\ref{fig:Fig_5}b. 

In order to elucidate the consequences of the inhomogenous magnetic field, let us assume that the magnetic field changes very slowly in space. Within this picture, which we will address as a local-density approximation regime and requires $|x-x_c|, \, \ell_B \ll \xi$, we can approximate the modified magnetic field with a constant $B(x)\approx B(x_0)$, where $x_0$ is the wavefunction center and it depends on the momentum as $x_0 = x_c -\zeta \ell_B^2 q_y$. As one moves away from the $K$ point, the LL wavefunctions is centered further away from $x_c$. From Eq.~\eqref{tiltedLL}, we know that the LL energy is proportional to $\sqrt{B(x_0)}$, which therefore translates into a decrease of the LL energies away from the $K$ point. As a result, the impact of inhomogeneous magnetic field corresponds to a deformation of the LLs which bend down away from the $K$ point. To be more quantitative, let us expand the Gaussian profile to the next order in $(x-x_c)/\xi$, yielding
\be\label{quartic_U0}
\bar n_\uparrow^{\text{(4)}}(x) = \frac{3Na}{4\sqrt{2\pi} \xi} \left[1-\frac{(x-x_c)^2}{2\xi^2} + \frac{(x-x_c)^4}{4 \xi^4} \right].
\ee
The resulting strain parameter is indeed inhomogeneous and reads
\begin{equation}\label{key}
\begin{split}
\tau^\text{(4)}_\text{eff}(x) =& \tau_\text{eff} \left[ 1 -\frac{7a^2 - 30a (x-x_c) +12 (x-x_c)^2}{4\xi^2}\right].
\end{split}
\end{equation}
where $\tau_\text{eff}$ is given by Eq.~\eqref{tau_U0}. Using the local density approximation, we therefore replace $x \ra x_0 = x_c -\zeta \ell_B^2 q_y$ and obtain the modified LLs energy levels
\begin{equation}\label{key}
\begin{split}
\epsilon^\text{(4)}_\nu =& \pm t \sqrt{\frac \nu 2  \tau_\text{eff}(1-\zeta q_y a)}\ \times \\
&\sqrt{1-\frac{7a^2}{4\xi^2}-\frac{15\zeta \ell_B^2}{2\xi^2} q_y a - \frac{3 \ell_B^4}{a^2\xi^2} (q_ya)^2}\,,
\end{split}
\end{equation}
which are shown in Fig.~\ref{fig:Fig_8}. The modified dispersion captures reasonably well the behavior of the lowest LLs, despite the fact that for this choice of parameters the LL wavefunctions are too broad as compared to $\xi$, thus weakening the validity condition of the local density approximation. 

\section{Numerical validation}
\label{sec:validation}

\begin{figure}[!b]
	\includegraphics[width=0.5\textwidth]{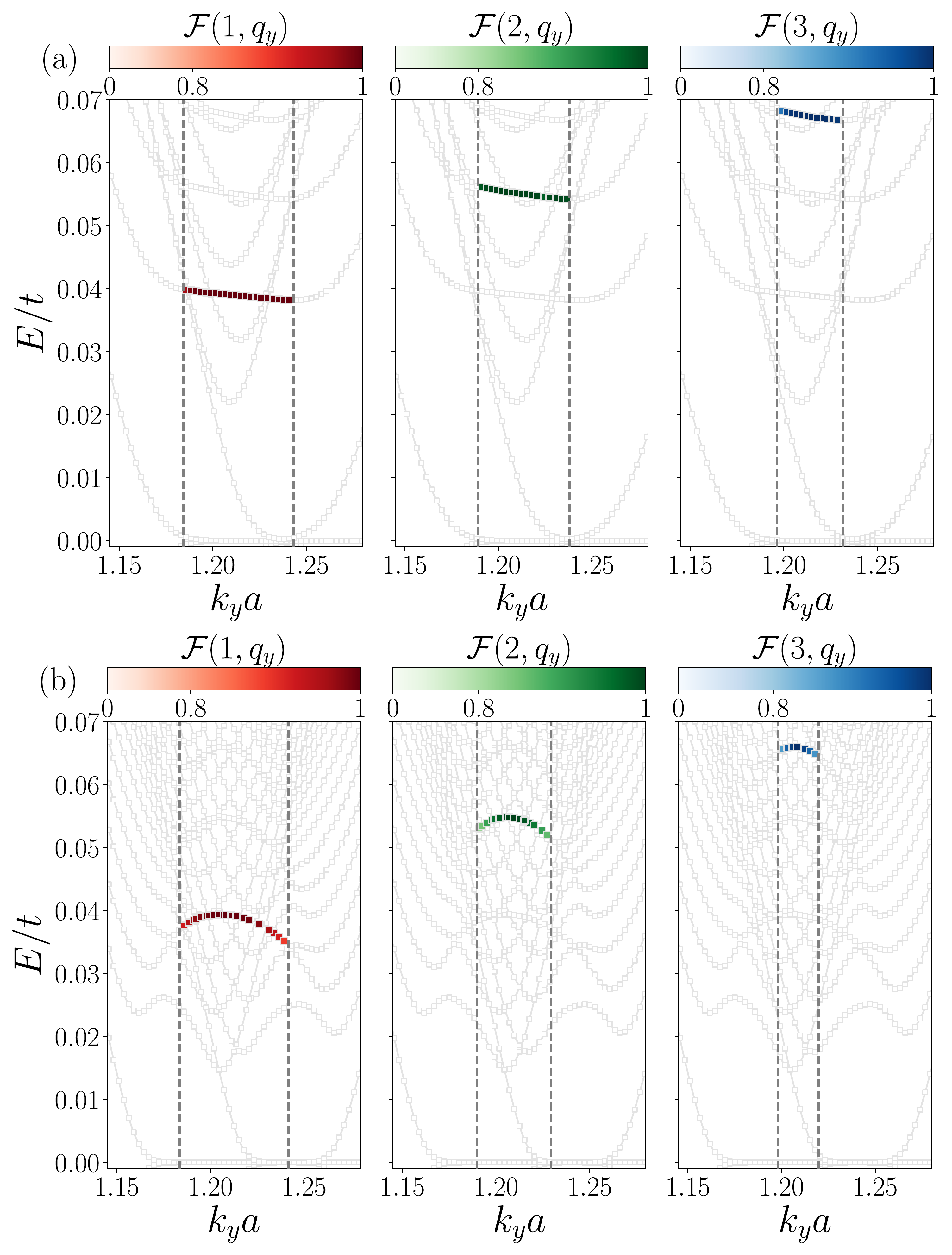}
	\caption{Spectra of $\hat H^\text{eff}_\downarrow$ obtained (a) in the TF regime and (b) in the noninteracting regime, where we superimpose the fidelity with the LL eigenstates computed in a range of momenta delimited by the vertical dashed lines. Each panel addresses the fidelity with a single Landau level. The parameters for the two regimes are chosen as in the previous plots. 
	}
	\label{fig:Fig_9}
\end{figure}

In this section, we further provide a numerical analysis of the results presented so far by showing a direct comparison of the eigenstates of $\hat H_{\downarrow}^\text{eff}$ with the ones expected from the ideal linear strain regime described by the Hamiltonian $\hat H_0$ and from the LLs description. In order to establish this comparison, we compute the fidelity $\mathcal F(\nu,q_y) = |\braket{\phi_{\nu,q_y}^{} |\psi_{q_y}^\downarrow}|^2$ where $\psi_{q_y}^\downarrow$ denotes the eigenstates of $\hat H_{\downarrow}^\text{eff}$ and $\phi_{\nu,q_y}^{}$ those of $\hat H_{0}$ corresponding to the $\nu^{\text{th}}$ LL. As mentioned in the previous section, the spectrum of $\hat H_{\downarrow}^{\text{eff}}$ differs from the one of $\hat H_{0}$ due to regions without strain or with magnetic field inhomogeneities. To minimize these effects, we focus on a certain window in momentum space $q_y \in [q_y^\text{min},q_y^\text{max}]$ centered around the $K$ point where we find high values of fidelity ($\mathcal F(\nu,q_y)\geq 0.8$) for bulk states in each $\nu^{\text{th}}$ LL. The results are shown in Fig.~\ref{fig:Fig_9}a,b for $\nu = 1,2,3$ in the TF and the non-interacting regimes. The parameters are chosen as in the previous section. 

In both cases, we observe that the eigenstates of $\hat H^\text{eff}_\downarrow$ reach a high fidelity with the ones of $\hat H_0$ near the center of the LL, namely near the $K$ point. However, in the non-interacting regime (Fig.~\ref{fig:Fig_9}b), the fidelity decreases as we go away from the $K$ point. This effect originates from the inhomogeneous magnetic field that modifies the wavefunction as compared to the expected ideal LL results. In particular, the magnetic length becomes space dependent and it increases when the magnetic field decreases, thus enlarging the tail of the wavefunctions and therefore lowering the fidelity. 

\begin{figure}[!t]
	\includegraphics[width=\columnwidth]{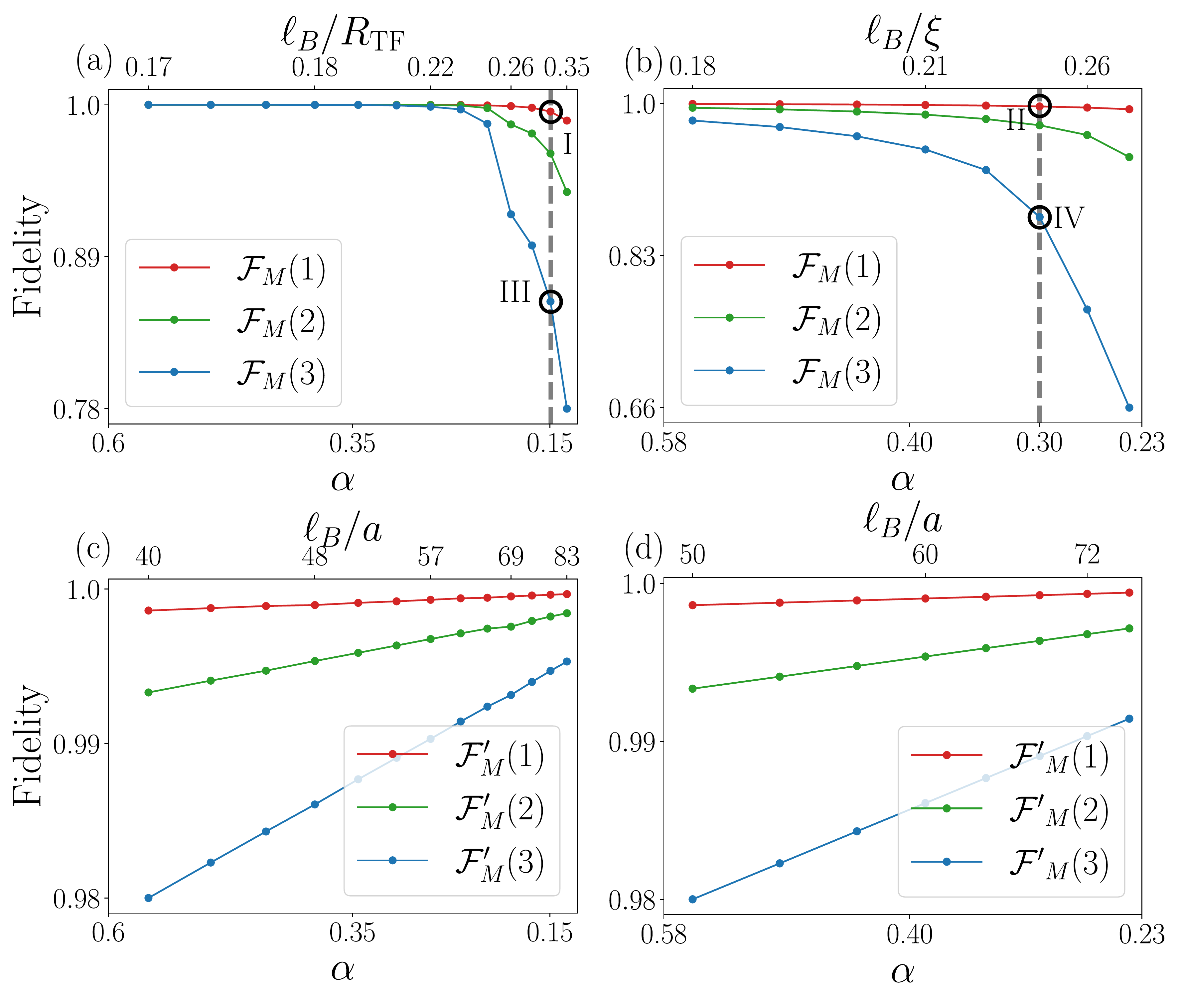}
	\caption{(a), (b) Fidelity $\mathcal F_M$ as a function of $\alpha$ ($\ell_B$) for (a) $U = 10^{-4} J$, $V_x = 10^{-6}J/a^2$, $N^\ua = 1.2 \times 10^5$ and (b) for $U = 0$, $V_x = 5.8\times 10^{-11}J/a^2$, $N ^\ua= 4.8 \times 10^5$. (c), (d) Fidelity $\mathcal F_M'$ as a function of $\alpha$ $(\ell_B)$ for the same parameters as in (a) and (b), respectively. The wavefunctions of the dots indicated by roman numbers in (a) and (b) are shown in Fig.~\ref{fig:Fig_11}. 
	}
	\label{fig:Fig_10}
\end{figure}

We now discuss how the LL picture correctly describes the results when we change the effective magnetic field, namely the parameter $\alpha$. In particular, we focus on the highest fidelity, denoted by $\mathcal F_M(\nu) = \max_{q_y}\varphi(\nu,q_y)$, as a function of $\alpha$. The results are shown in Fig.~\ref{fig:Fig_10}a for the TF regime and in Fig.~\ref{fig:Fig_10}b for the non-interacting regime. The content of these plots can be understood through a lengthscale analysis. The smallest lengthscale is the lattice spacing $a$, whereas the largest one (besides  the size of the system $L_x$) is related to the BEC size, namely $\xi$ and $R_{\text{TF}}$ for the non-interacting and the TF regimes, respectively. The relevant lengthscale for LLs physics is the magnetic length $\ell_B$. We therefore conclude that the ideal situation to observe LLs requires $a \ll \ell_B\ll \xi,R_\text{TF}$. 

In both regimes, we find that the fidelity $\mathcal F_M$ is close to 1 for large values of $\alpha$ whereas it drops as $\alpha$ decreases, see Figs.~\ref{fig:Fig_10}a,b. The best scenario is therefore reached for sufficiently large values of $\alpha$. However, when $\alpha$ becomes too large the LL pictures breaks down since the magnetic length $\ell_B$ becomes smaller and lattice spacing effects take place. We can already see this trend for the values of $\alpha$ chosen in this analysis, as shown in Figs.~\ref{fig:Fig_10}c,d. We indeed observe that the fidelity $\mathcal F'_M = \max_{q_y} |\braket{\phi_{\nu,q_y}| \Psi_{\nu,q_y}}|^2$ between the analytical relativistic Landau levels $\Psi_{\nu,q_y}$ given by Eq.~\eqref{Landaustates} and $\phi_{\nu,q_y}$, decreases as $\alpha$ increases. The impact of the lattice discreteness is more effective for higher LLs, which have more nodes and thus a less smooth wavefunction, which results in a lower fidelity. A second reason for the drop in fidelity $\mathcal F'_M$ as $\alpha$ increases comes from the asymmetry (or parity breaking) with respect to the center $x_c$ that the wavefunctions manifest and that can be identified by inspecting Fig.~\ref{fig:Fig_3}. There one can recognize that the left peaks of the wavefunctions have different heights with respect to the right peaks, whereas the analytical LL states do not. This feature, which was already noticed in Ref.~\cite{Salerno2016} and caused by terms that have been neglected in the effective Dirac description, is negligible for small strain values but becomes more and more relevant for larger ones, thus causing a distinct mechanism for a mismatch with the ideal LL wavefunctions and the drop in fidelity. 

\begin{figure}[!b]
	\includegraphics[width=\columnwidth]{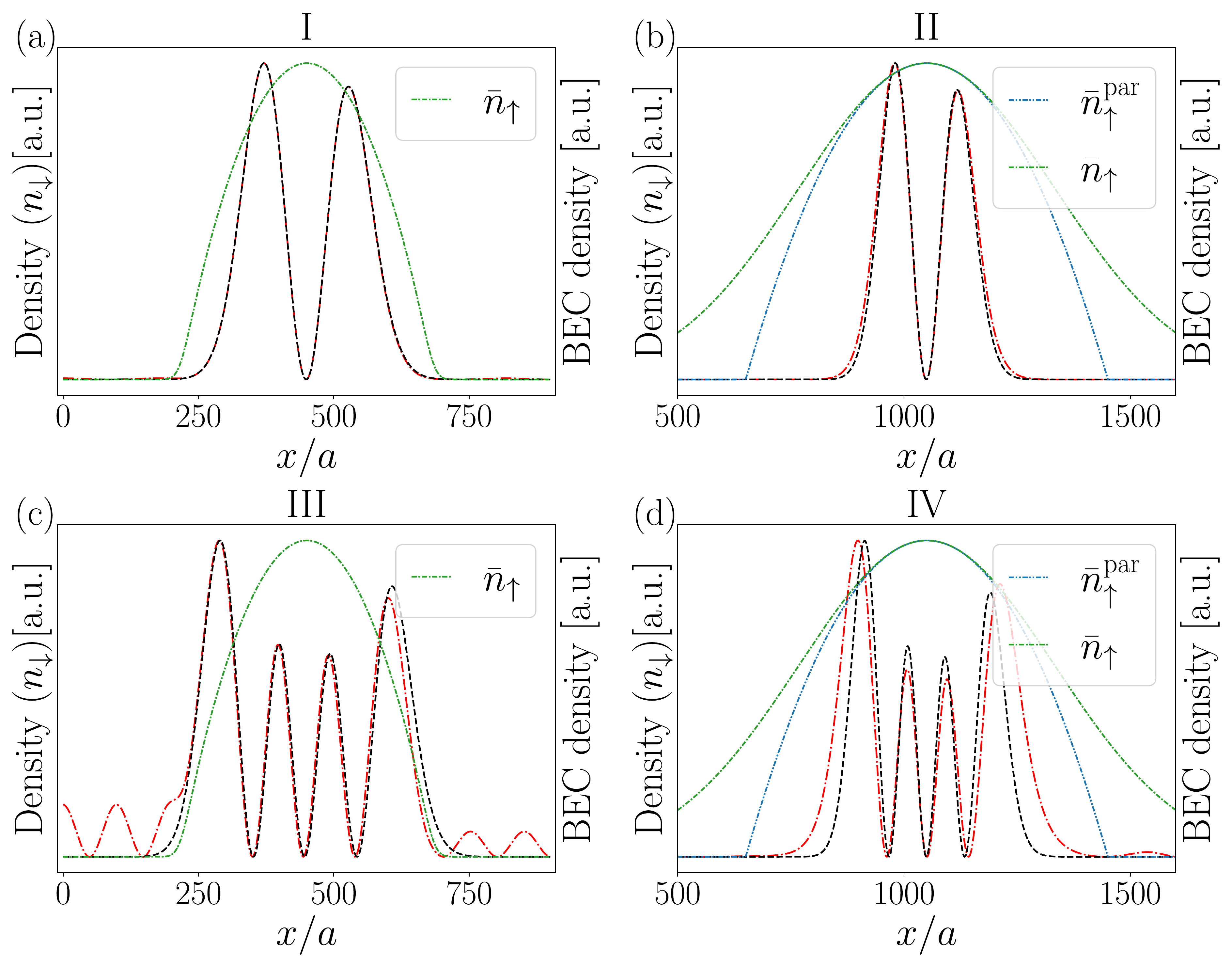}
	\caption{Density from the eigenvectors of (long dashed-dotted) $\hat H^\text{eff}_{\downarrow}$ and (dashed) $\hat H_0$ corresponding to the circles indicated in Figs.~\ref{fig:Fig_10}a,b for (a), (b) $q_y a=0$ and (c), (d) $q_y a=-0.00067$. In all panels, we plot only the $A$ component.}
	\label{fig:Fig_11}
\end{figure}

We can therefore conclude that the ideal regime requires a not so large value of $\alpha$ because novel effects that invalidate the LL picture take place, as the condition $\ell_B \gg a$ is not satisfied anymore. On the other side, when $\alpha$ decreases, the wavefunction broadens and the condition $\ell_B \ll \xi, R_{\text{TF}}$ breaks down. We may encounter situations where the lowest LL has a very good fidelity (see Figs.~\ref{fig:Fig_11}a,b) when centered near $x_c$ whereas the highest LLs are more strongly affected given their larger size (Figs.~\ref{fig:Fig_11}c,d). In the TF regime, the wavefunctions can indeed cross the interface and hybridize with the planewave solutions of the unstrained region. This effect is shown in Fig.~\ref{fig:Fig_11}c. In the non-interacting regime, one must instead consider the intermediate region where the BEC density is non-parabolic. In this region, the magnetic field is nonuniform as we discussed before, thus implying that the magnetic length acquires a space-dependence and becomes larger as we go away from the center, which in turn broadens the wavefunction, as shown in Figs.~\ref{fig:Fig_11}d. 

\section{Experimental realization and probing}\label{sec:realization}

In this section, we outline a method to experimentally implement the model discussed in the previous section by using a time-dependent scheme and we then discuss possible detection protocols.

\subsection{Floquet scheme}

In order to generate the correlated hopping parameters given in Eq.~\eqref{effective_tj}, we combine a Floquet engineering method inspired from Ref.~\cite{DiLiberto2014} where interactions are modulated in time, and we combine it with the resonant driving scheme analyzed in Ref.~\cite{Goldman2015} for a double-well system. Before discussing the coupling between the $\uparrow$ and $\downarrow$ species, let us briefly review how a resonant Floquet driving scheme can be implemented to engineer tunneling amplitudes in a double-well system. Let us consider a single species ($\downarrow$) described by the time ($\tau$)-dependent Hamiltonian $\hat H(\tau) = \hat H_{\text{hop}} + \hat{\mathcal V} (\tau)$, where
 \begin{equation}\label{key}
 \begin{split}
 \hat H_{\text{hop}} &= -t \hat c^{\dagger}_{\downarrow,0} \hat c_{\downarrow,1} - t\hat c^{\dagger}_{\downarrow,1} \hat c_{\downarrow,0} + \Delta \hat n_{\downarrow,1}\\
 \hat{\mathcal V}(\tau) &=\mathcal K_0 \cos(\Omega \tau) \hat n_{\downarrow,0} +\mathcal K_1\cos(\Omega \tau) \hat n_{\downarrow,1},
 \end{split}
 \end{equation}
with $\Omega$ the modulation frequency, $\hat c^\dagger_{\downarrow,i}$ ($\hat c^{}_{\downarrow,i}$) the creation (annihilation) operator of the $\downarrow$ atoms at site $i \in \{0,1\}$ and $\hat n_{\downarrow,i} \equiv \hat c^\dagger_{\downarrow,i} \hat c^{}_{\downarrow,i}$. The parameter $\Delta$ describes an energy off-set between the two sites. Differently from the model in Ref.~\cite{Goldman2015}, we have imposed a time modulation for both sites. If the resonant condition $\Delta = \hbar \Omega \gg J$ is met, we can follow the standard procedure of changing basis to the rotating frame through the unitary transformation $\hat{\mathcal{R}} = \hat{\mathcal{R}}_1 \hat{\mathcal{R}}_2$, where
\begin{equation}\label{key}
\begin{split}
\hat{\mathcal{R}}_1(\tau) &= e^{i\Omega \tau \hat n_{\downarrow,1 } },\\
\hat{\mathcal{R}}_2(\tau ) &= e^{i\left[\frac{\mathcal K_0}{\hbar \Omega} \sin(\Omega \tau) \hat n_{\downarrow,0} + \frac{\mathcal K_1}{\hbar \Omega} \sin(\Omega \tau)\hat n_{\downarrow,1}\right]}.
\end{split}
\end{equation}
The resulting Hamiltonian transforms into
\begin{equation}\label{key}
\begin{split}
\hat{\mathcal H}(\tau) &=\hat{\mathcal{R}}\hat H(\tau)\hat{\mathcal{R}}^\dagger - i\hbar \hat{\mathcal{R}} \partial_\tau \hat{\mathcal{R}}^\dagger
\\ &= -t e^{i(\mathcal K_0-\mathcal K_1)\sin(\Omega \tau)/\hbar \Omega - i\Omega \tau} \hat c^\dagger_{\downarrow,0} \hat c^{}_{\downarrow,1} + \text{h.c.}
\end{split}
\end{equation}
We obtain an effective time-independent Hamiltonian by taking the time-average of $\hat{\mathcal H}(\tau)$ which reads
\begin{equation}\label{key}
\begin{split}
\hat{\mathcal H}^{\text{eff}} = -t^{\text{eff}} \hat c^\dagger_{\downarrow,0} \hat c^{}_{\downarrow,1} + h.c.,
\end{split}
\end{equation}
where $t^{\text{eff}} = t \mathcal J_1((\mathcal K_0-\mathcal K_1)/\hbar \Omega)$, with $\mathcal J_1(x)$ being the first Bessel function of the first kind. When its argument is much smaller than 1, namely $\mathcal{K}_0-\mathcal K_1\ll \hbar \Omega$, it can be linearised as $\mathcal J_1\left(x\right) \approx x / 2$
thus yielding an effective hopping amplitude
\begin{equation}\label{key}
t^{\text{eff}} \approx t\frac{ \mathcal K_0 - \mathcal K_1}{2\hbar \Omega}.
\end{equation}

In order to understand how to generate the correlated-hopping term, let us replace $\mathcal V(\tau)$ by an interaction term between the $\downarrow$ species and the $\uparrow$ species that reads
\begin{equation}\label{key}
\hat H_{\uparrow \downarrow} = U_{\uparrow\downarrow}(\tau) (\hat n_{\uparrow,0} \hat n_{\downarrow,0} + \hat n_{\uparrow,1} \hat n_{\downarrow,1}),
\end{equation}
where $U_{\uparrow\downarrow}(\tau) \equiv \mathcal U \cos(\Omega \tau)$, as in Ref.~\cite{DiLiberto2014}. 
We can then identify $\mathcal K_i = \mathcal U \hat n_{\uparrow,i}$ and apply the resonant driving scheme described before, which yields 
\begin{equation}\label{teff_final}
t^\text{eff} = t \frac{\alpha}{3} (\hat n_{\uparrow,0} - \hat n_{\uparrow,1}),
\end{equation}
with $\alpha \equiv 3\mathcal U/2\hbar\Omega$. We immediately conclude that the following  condition $\alpha |\braket{\{n_{\uparrow,i}'\}|(\hat n_{\uparrow,0} \!-\!\hat n_{\uparrow,1})|\{n_{\uparrow,i}\} }| \! \ll \! 1$, where $\{n_{\uparrow,i}\}$ labels the Fock states, must be satisfied in order to linearize the Bessel function. 

The scheme that we have discussed so far provides the correlated-hopping term that is central to our model of strain in Sec.~\ref{Model}. However, the $\downarrow$ atoms must also possess a dominant tunneling amplitude that is unaffected by the interaction with the $\uparrow$ atoms. In order to guarantee such a process, we assume that the $\downarrow$ atoms are prepared in a superposition of two hyperfine states $\ket \downarrow \equiv (\ket + + \ket -)/\sqrt 2$. Each component must experience the same optical lattice, and we indicate the corresponding tunneling amplitudes as $t^+$ and $t^-$. However, we only let the $\ket -$ component interact with the $\uparrow$ atoms. In this configuration, the hopping amplitude of the $\ket -$ component will be renormalized as in Eq.~\eqref{teff_final}, while the hopping amplitude of the $\ket+$ component remains unaffected. The total hopping amplitude for a $\downarrow$ atom in such a superposition will therefore be given by
\begin{equation}\label{key}
t^{\text{eff}} = (t^+ + \kappa t^-)/2,\quad \kappa =   \mathcal J_1(2\alpha(\hat n_{\uparrow,0}- \hat n_{\uparrow,1})/3).
\end{equation}

\begin{figure}[!t]
	\includegraphics[width=0.4\textwidth]{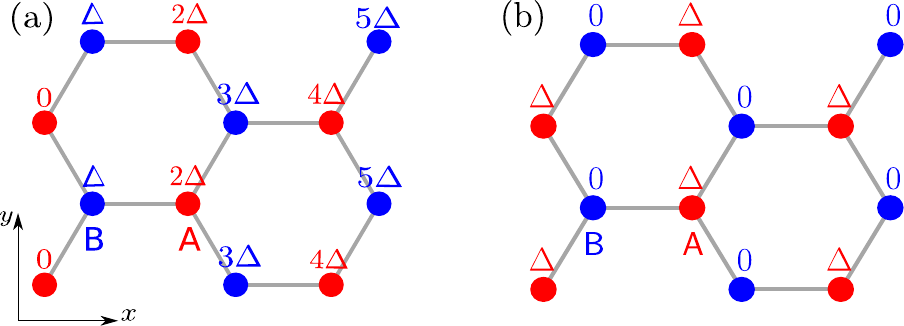}
	\caption{Two possible Floquet engineering schemes of strain as described in the main text. (a) Energy offset $\Delta$ linearly growing in the $x$ direction. (b) Homogeneous energy offsets $\Delta$ only on the $A$ sites.}
	\label{fig:Fig_12}
\end{figure}

The double-well scheme that we have discussed must be applied to an extended honeycomb lattice in order to reproduce uniaxial strain. We have identified two possible implementations, which we sketch in Fig.~\ref{fig:Fig_12}. The first scheme consists in applying an energy offset on each lattice site that grows  along the $x$ axis. The Floquet results presented for the double-well case apply here to the hopping between two neighboring sites of the honeycomb lattice after identifying site 1 as the one with the highest energy offset and site 0 as the one with the lowest energy offset. The generalization of Eq.~\eqref{teff_final} to the honeycomb lattice is therefore
\begin{equation}\label{}
t^{\text{eff}}_j = t \frac{\alpha}{3} \sigma_j  (\hat n_{\uparrow,\mathbf r + \boldsymbol \delta_j}-\hat n_{\uparrow,\mathbf{r}}) ,\quad \mathbf r \in \mathcal A
\end{equation}
with $\sigma_1 = 1$ and $\sigma_{2,3} = -1$, which is exactly what we studied in the previous sections. 

The second scheme presents an offset ($\Delta$) only for the $\mathcal A$ sublattice and none for $\mathcal B$ sublattice. In the double-well representation, this means that the site $0$ is always a $B$ site and the site $1$ is always an $A$ site. The effective hopping amplitude in this case reads
\begin{equation}\label{key}
t^{\text{eff}}_j = t \frac{\alpha}{3} (\hat n_{\uparrow,\mathbf r }-\hat n_{\uparrow,\mathbf{r}+ \boldsymbol \delta_j}),\quad \mathbf r \in \mathcal A.
\end{equation}
Therefore, by following the same reasoning as the one leading to Eq.~\eqref{A_corr}, the vector potential $\mathbf A$ for the present configuration is given by
\begin{equation}\label{key}
\mathbf A = \left(0,\xi \frac{\hbar \tau_{\text{eff}} }{3a^2} (x-x_c) - \frac{\zeta\hbar\tau}{12a}\right),
\end{equation}
which corresponds to a magnetic field larger by a factor of $3$. The corresponding LLs energies are
\be\label{tiltedLL2}
\epsilon^{\text{LL}}_{\nu} (q_y) = \pm t\sqrt{\nu \frac{3\tau_\text{eff}}{2}} \sqrt{1-\xi q_y a},  \qquad \nu \in \mathbb N,
\ee
For $\nu = 1$, the energy gap is thus larger by a factor $\sqrt 3$. As this scheme provides a larger gap, it would be more suitable for the experimental detection of LL physics.

\subsection{Further effects and probing methods}

In the discussion above, we have not considered the possible renormalization of the hopping parameter $J$ of the $\uparrow$ atoms due to the time modulation of interactions. As presented in Ref.~\cite{DiLiberto2014}, if no energy offset is experienced by the $\ua$ atoms and if $\Omega \gg J,\mathcal U$ then the hopping amplitude for the $\ua$ bosonic atoms is renormalized as $J \to J_{\text{eff}} = J\mathcal J_0(\alpha (\hat n_{\downarrow}^0 - \hat n^1_{\downarrow}) )$. When the argument of the Bessel function is sufficiently small, the hopping amplitude is unaffected and no back action of the $\da$ atoms takes place onto the $\ua$ atoms. This condition can be met when the density of $\da$ atoms is small or when it is homogeneous. So far, we have not specified the statistics of $\da$ atoms. If we consider a Fermi gas of $\da$ atoms, we may have both conditions met at once because Pauli exclusion principle will prevent the onsite density to exceed one atom per site and it  will also broaden the density distribution in the presence of a harmonic trap. The latter may result in a very flat density profile over the range occupied by the BEC of $\ua$ atoms. If we consider a gas of bosonic $\da$ atoms, we will instead have to enforce a low density or a homogeneous distribution in order to prevent the back action on the $\ua$ atoms. Despite the arguments presented here to neglect back action effects, it would nevertheless be interesting to include those, as they will provide a distinct opportunity to enrich the strain picture discussed in this work with dynamical effects.

A separate discussion for the actual implementation of the model concerns the trapping potential. The strain model that we have analyzed requires a strongly anisotropic harmonic trap experienced by the BEC of $\ua$ atoms with $\om_x \gg \om_y$, where $\om_{x,y}$ are the harmonic trapping frequencies in the two spatial directions. In this work we have actually considered $\om_y \ra 0$ to simplify the theoretical analysis. However, we have not included trapping effects on the $\da$ atoms, assuming that one can independently control the confinement of the two atomic species. In this case, several scenarios are possible, which affect the back action discussed before and the corresponding probing methods. 
When the harmonic trap is absent or negligible, we obtain the picture discussed in this work for the single-particle spectrum. However, in the presence of a strong confinement and for fermionic $\da$ atoms, we will have the opportunity to reveal the presence of LL physics when the corresponding Fermi level at the center of the system is fixed between the LL gaps. In this case, LLs will manifest through jumps in the density profile that will confer the typical wedding cake structure to the fermionic gas, see Ref.~\cite{Pekker2015}. Near these jumps, which are going to be partly smeared out because the LLs are not perfectly flat, we also expect to find valley-dependent edge modes that represent a clear signature of the valley Hall physics.

In order to directly probe the properties of LLs, there are several available techniques that can be employed. One possibility is to monitor the real space dynamics of a wave packet (either for fermions or for bosons) of $\downarrow$ atoms near the $K$ point, which will exhibit a cyclotron motion as shown in Ref.~\cite{Pekker2015}. Another possibility applies to a uniform fermionic gas of $\downarrow$ atoms at half-filling. In this case, circular lattice shaking will allow to spectroscopically resolve the LLs by measuring the absorbed energy. Moreover, band mapping techniques make possible to identify valley dependent absorption processes, thus allowing to extract the corresponding valley dichroism \cite{Asteria2019}. 

\section{Conclusions}
\label{sec:conclusions}

In this work, we have presented a strategy to generate a strain field in optical lattices that is implemented by coupling an atomic species to a trapped BEC via well-tailored density-assisted tunneling terms. By changing the shape of the BEC profile or the type of density-assisted tunneling terms, generic strain profiles can in principle be generated. We have focused on the implementation of uniaxial linear strain applied along one of the three crystalline axes of the lattice, which is realized by considering a strongly-anisotropic harmonic trapping potential. We have then discussed two limits of interest, namely the non-interacting and the Thomas-Fermi limits. After investigating the spectral features, we have identified the Thomas-Fermi regime as most suitable to reproduce the ideal linear strain configuration. Indeed, this regime minimizes the effects of regions with inhomogeneous magnetic field and requires smaller atomic clouds. At the same time, the Thomas-Fermi regime may also provide an interesting scenario to study the effect of quantum fluctuations originating from the phonon modes of the BEC or the effect of exciting the collective modes of the BEC, which would provide time-dependence to the synthetic gauge field. Some of these effects have a correspondence in solid-state system and originate from the lattice vibrations of the crystal. An other interesting scenario which is more specific to cold atoms is to investigate the strongly-interacting regime for the bosonic gas near the Mott insulator phase. In this case, low filling and strong quantum fluctuations would provide a very different regime as compared to what is studied in solid state materials. Our results therefore suggest a distinct direction to investigate the interplay of dynamical gauge fields, as realized through synthetic strain fields, and quantum matter with ultracold atoms.

\section{Acknowledgements}

We would like to thank G. Salerno and C. Schoonen for fruitful discussions. Work in Bruxelles is supported by the ERC Starting Grant TopoCold and the Fonds De La Recherche Scientifique (FRS-FNRS, Belgium) and the Universit\'e Libre de Bruxelles. Work in Innsbruck is supported by the QuantERA grant MAQS via the Austrian Science Fund FWF No I4391-N.

\bibliography{biblio}

\end{document}